\documentclass[onecolumn]{aa}

\usepackage{xcolor}
\usepackage{graphicx}
\usepackage{epsfig}
\usepackage[varg]{txfonts}
\usepackage{graphicx}
\usepackage{amsmath}
\usepackage[english]{babel}
\usepackage[titletoc,title]{appendix}
\usepackage{txfonts}
\usepackage{times}
\usepackage{natbib}
\usepackage{cancel}

\begin{document}

\title{Outflows in the Gaseous Discs of Active Galaxies and their impact on Black Hole Scaling Relations} 

\author{N. Menci$^1$, F. Fiore $^{2}$, F. Shankar$^{3}$, L. Zanisi$^{3}$, C. Feruglio$^{2}$}
\institute{INAF - Osservatorio Astronomico di Roma, via Frascati 33, I-00078 Monte Porzio, Italy\and
INAF - Osservatorio Astronomico di Trieste, via G.B. Tiepolo 11, I-34131, Trieste, Italy \and 
School of Physics \& Astronomy, University of Southampton, Highfield, Southampton SO17 1BJ, UK}

\titlerunning{AGN Outflows and Black Hole Scaling Relations}
\authorrunning{Menci, Fiore, Shankar, Zanisi, Feruglio}

\abstract{To tackle the still unsolved and fundamental problem of the role of Active Galactic Nuclei (AGN) feedback in shaping galaxies, in this work we implement a new physical treatment of AGN-driven winds into our semi-analytic model of galaxy formation. To each galaxy in our model, we associate solutions for the outflow expansion and the mass outflow rates in different directions, depending on the AGN luminosity, on the circular velocity of the host halo, and on gas content of the considered galaxy. To each galaxy we also assign an effective radius derived from energy conservation during merger events, and a stellar velocity dispersion self-consistently computed via Jeans modelling. We derive all the main scaling relations between Black hole (BH) mass and total/bulge stellar mass, velocity dispersion, host halo dark matter mass, and star formation efficiency. We find that our improved AGN feedback mostly controls the dispersion around the relations but plays a subdominant role in shaping slopes and/or normalizations of the scaling relations. Including possible limited-resolution selection biases in the model provides better agreement with the available data. The model does not point to any more fundamental galactic property linked to BH mass, with velocity dispersion playing a similar role with respect to stellar mass, in tension with present data. In line with other independent studies carried out on comprehensive semi-analytic and hydrodynamic galaxy-BH evolution models, our current results signal either an inadequacy of present cosmological models of galaxy formation in fully reproducing the local scaling relations, in terms of both shape and residuals, and/or point to an incompleteness issue affecting the local sample of dynamically-measured BHs.}

\maketitle

\section{Introduction} 

Understanding the processes governing the  growth of supermassive Black Holes (BHs) constitutes a major step toward the building up of a theoretical framework for galaxy formation. In all current {\it ab-initio} galaxy formation models 
such a growth results from the interplay between the growth of dark matter (DM) halos, the collapse and the cooling of gas in the DM potential wells, the dynamical processes affecting the distribution of gas (merging, interaction, and gravitational instabilities), and the counteracting effects of outflows
(e.g., Granato et al. 2004; Menci et al. 2005; Hopkins et al. 2006; Lapi et al. 2006; Shankar et al. 2006; Guo et al. 2011; Menci et al. 2014; Fanidakis et al. 2012; Habouzit et al. 2021, and references therein). The latter are due both to the supernovae explosions following star formation (see, e.g., White \& Rees 1978; Dekel \& Silk 1986) and to the radiation field of Active Galactic Nuclei (AGN) powered by the process of accretion onto the central BHs. 

A putative co-evolution between central BHs and their host galaxies and dark matter halos is supported by the observational evidence of a series of correlations between the BH mass $M_{BH}$ and several host galaxy properties (see, e.g., Ferrarese \& Ford 2005; Kormendy \& Ho 2013; Graham 2016, for reviews). In particular, an evident
correlation between the mass of the BH $M_{BH}$ and the stellar mass of the host galaxy bulge $M_{*,b}$, with an intrinsic scatter of $\lesssim 0.4$ dex in BH mass at fixed bulge mass, has been several times reported in the literature (e.g., Magorrian et al. 1998; Marconi \& Hunt 2003; Kormendy \& Ho 2013; McConnell \& Ma 2013; L\"asker et al. 2014; Saglia et al. 2016; Davis, Graham \& Cameron 2018; de Nicola, Marconi \& Longo 2019; see, e.g., Graham 2016 for a review). 
A possibly even tighter correlation between $M_{BH}$ and stellar velocity dispersion $\sigma$, with an intrinsic scatter as low as $\lesssim 0.3$ dex (e.g., Gebhardt et al. 2000, but see more recent estimates by, e.g., Sahu et al. 2020), has often been interpreted as evidence of the key role played by AGN in self-regulating both BH growth and star formation in the host galaxy via some wind/jet-driven feedback mechanism (see reviews by, e.g., Shankar 2009; Alexander \& Hickox 2012; Somerville \& Dav{\'e} 2015), although galaxy-galaxy mergers could still preserve its tightness but possibly modify its shape (e.g., Graham 2022, and references therein). Analysis of the pairwise residual correlations applied to the BH-galaxy scaling relations have also revealed that stellar velocity dispersion $\sigma$ appears more correlated to central BH mass $M_{BH}$ than other galactic variables (e.g., Bernardi et al. 2007; Hopkins et al. 2007; Shankar et al. 2016, 2017, 2019; Iannella et al. 2021), which is a trend expected in AGN feedback-driven co-evolution models (e.g., Silk \& Rees 1998; King 2003).

When moving to even larger scales, central BHs continue to show evidence for correlations between the BH mass and the total dynamical mass of the host galaxy (Bandara et al. 2009), with the number of globular clusters (Burkert \& Tremaine 2010), the galactic rotation velocity $v_{rot}$ (Davis, Graham \& Combes 2019;  Robinson et al. 2021), and the mass of the entire host DM halo mass (e.g., Ferrarese 2002; Baes et al. 2003; Pizzella et al. 2005; Volonteri, Natarajan, Gultekin 2011). A clear link between the BH mass and host dark matter halos is also evident from the clustering strength of AGN, at least at $z\lesssim 0.5$ (e.g., Shankar et al. 2020; Allevato et al. 2021; Powell et al. 2022). A correlation between $M_{BH}$ and $M_h$ has also been measured by  Marasco et al. (2021) who considered both early- and late-type systems for which $M_h$ is determined either from globular cluster dynamics or from spatially resolved rotation curves. It is of course possible that correlations between BHs and the larger-scale systems such as the host halos could be secondary ones, ultimately induced by the underlying correlation between galaxy mass/velocity dispersion and host halo mass (e.g., Ferrarese 2002).

It is the aim of this paper to explore the predicted normalisation, shape, scatter, and even pairwise residuals of the main scaling relations between BH mass and host galaxy properties, namely stellar mass, stellar velocity dispersion, and the mass of the host dark matter halo, with the goal of dissecting how the above correlations arise from the complex interplay between DM halo growth, the inflow of gas, galaxy mergers, and the stellar-AGN feedback mechanisms at play in galaxy evolution models. Shedding light on the origin of the BH-host scaling relations has constituted a major goal of galaxy formation models in the last decades, and possibly all present-day cosmological models now include more or less refined recipes to self-consistently grow massive BHs at the centre of galaxies, as also mentioned above (e.g., Cirasuolo et al. 2005; Fontanot et al. 2015, 2020; Habouzit et al. 2021, and references therein). However, our present work differs from previous attempts in two main respects: i) it adapts a well-tested and realistic kinetic, ``blast-wave'' AGN feedback recipe into a comprehensive semi-analytic model; ii) it self-consistently explore the possible impact of selections biases whilst analysing shapes and residuals in the BH-galaxy scaling relations. We further discuss these points below.

A critical aspect in realistic models of the co-evolution of BHs and their hosts is constituted by the modeling of AGN feedback. In fact, while the role played by the growth of DM haloes and the ensuing inflow and cooling of gas in the building of massive BHs is relatively well established in galaxy formation models and simulations (see, e.g., Somerville \& Dave' 2015;  Naab \& Ostriker 2017 for a review), large uncertainties still affect the implementation of reliable descriptions of AGN feedback, constituting a key self-regulation mechanism in the process of BH growth (see, e.g., Alexander \& Hickox 2012 for a review).  Cosmological simulations  adopt various different phenomenological approximations to represent AGN feedback, often including multiple feedback ``modes'' associated with rapid (close to Eddington) or slow (very sub-Eddington) accretion onto the central BH (see Somerville \& Dav\'e 2015, and references therein for reviews). The large dispersion in the predicted AGN luminosity functions from different simulations presumably largely reflects the major uncertainties in modelling these processes. On the other hand, semi-analytic models often adopt a parametric, phenomenological description  of AGN feedback, and compute the amount of ejected gas on the basis of parametric laws based on energy-balance arguments, and do not consider the complex two-dimensional structure of the outflows. 
While an approach based on a physical model for the expansion of AGN-driven shocks in the galactic gas has been adopted in Menci et al. (2008), this was still based on a isotropic model for shock expansion. However, to act as 
an effective negative feedback on star formation and BH accretion, AGN winds must couple to and affect cold galactic gas which is rotationally supported and settled in a disc. The large density of the gas distribution in the direction parallel to the plane of the disc strongly inhibits the shock expansion in such a direction, while the gas is preferentially ejected perpendicular to the disc, resulting in an overall fraction of ejected interstellar medium  lower than in one-dimensional (isotropic)  models 
(see, e.g., Faucher-Giguere \& Quataert 2012; Hartwick, Volonteri, Dashyan 2018). 
Moreover, all treatments of AGN feedback adopted so far in galaxy formation models have not been tested in detail against the observed properties of AGN winds, which are being increasingly studied in the recent years. Indeed, since  the works by Cicone et al. (2014) and Fiore et al. (2017), samples with more than a hundred outflow measurements have been assembled, with detected massive winds at different scales (sub-pc to kpc) and with different molecular/ion compositions. In a previous paper (Menci et al. 2019), we computed the two-dimensional expansion of outflows driven by AGN  in galactic discs as a function of the global properties of the host galaxy and of the luminosity of the central AGN. We derived the expansion rate, the mass outflow rate, and the density and temperature of the shocked shell in the case of an exponential profile for the disc gas, for different expansion directions $\theta$  with respect to the plane of the disc. Having expressed our model results in terms of global properties of the host galaxies, we compared our predictions to a large sample of 19 outflows (mostly molecular, except for one object) in galaxies with measured AGN luminosity and gas mass, and with estimated total mass. This allowed us to test the model through a detailed, one-by-one comparison with the data. 
The two-dimensional structure that we obtain for the outflows is characterized by a shock expansion that follows the paths of least resistance with an elongated shock front in the direction perpendicular to the disc. The larger outflow velocities attained in the  direction perpendicular to the disc easily exceed the escape velocity at the virial radius in a short timescale, while the slower expansion of the shock in the plane of the disc can prevent the escape of gas in this direction within the lifetime of the AGN. The overall ejected gas fraction differs substantially from that obtained in spherical models, both in the total value and in its dependence on the galaxy properties. Here we implement such a two-dimensional description in our semi-analytic model of galaxy formation. For each model galaxy in our Monte Carlo realizations, we associate the outflow solutions corresponding to its properties (gas mass, total DM mass, AGN luminosity), and derive the opening angle  generated by the outflows and the total ejected gas mass. This   improved and observationally-tested description of AGN feedback constitutes a novel upgrade of our galaxy formation model that allows us to compute the scaling relations between the BH mass and the different galaxy properties on more physically motivated grounds. In addition, we have improved our model by implementing the computation of  the size $R_e$ of  the bulge component of the model galaxies, and then by deriving the bulge stellar velocity dispersion $\sigma$. This is computed by solving the spherical Jeans equation for each model galaxy. This allows us not only to predict - among other observables - the $M_{BH}-\sigma$ relation, but also to derive for each model galaxy the BH sphere of influence and hence to investigate the effects of angular resolution-related selection effects. 
  
In this work we indeed also explore the impact of possible observational biases that may affect the local sample of dynamically measured supermassive BHs, and thus in turn affect the comparison between theoretical predictions and the observed correlations, since the latter may differ from the intrinsic relations.
Shankar et al. (2016; 2017; 2019), following in the footsteps of other groups (e.g., Batcheldor 2010, Morabito \& Dai 2012), have put forward the hypothesis that local samples of quiescent galaxies with dynamically measured $M_{BH}$ may suffer from an angular resolution-related selection effect, which could bias the observed scaling relations between $M_{BH}$ and host galaxy properties away from the intrinsic relations. In fact, insufficient resolution could prevent reliable BH mass estimates whenever the spatial resolution $r_{crit}$ is larger than the BH sphere of influence $r_{infl}\equiv G\,M_{BH}/\sigma^{2}$. Specifically, the condition $r_{infl}\geq r_{crit}$ may lead to a sample biased towards larger masses of dynamically measured $M_{BH}$, at fixed galaxy stellar mass or stellar velocity dispersion.
The above authors investigated  the effect of such a bias assuming different intrinsic relations as an input for aimed Monte Carlo simulations, concluding that  the $M_{BH}-M_*$  relation is more strongly biased than the $M_{BH}-\sigma$  relation. Since such a bias only affects dynamical measurements of $M_{BH}$, this could explain the difference in the scaling relations  obtained for samples of active AGN (Ho \& Kim 2014; Martın-Navarro \& Mezcua 2018; van den Bosch 2016;  Greene et al. 2016;  Busch et al. 2014, Reines \& Volonteri 2015; Bentz \& Manne-Nicholas 2018). In fact, in such samples $M_{BH}$ is not determined from  spatially resolved dynamical measurements, but rather  from the assumed virial motions of the gas in the broad line region (BLR) orbiting in the vicinity of the central BH. However,  such measurements are affected by the uncertainty related to the 
kinematics, geometry, and inclination of the BLR clouds (e.g., Ho \& Kim 2014, and references therein), usually parametrised in terms of a free parameter $f_{vir}$ relating the measured velocity of the clouds $\Delta V$ and its size $r$ to the BH mass $M_{BH}=f_{vir}\,r\, (\Delta V)^{2}/G$ (but see also, e.g., Marculewicz \& Nikolajuk 2020 on how to adapt the virial factor $f_{vir}$ based on the spectral properties of the quasars). Although the degree and extent of a resolution bias in the current sample of dynamically measured supermassive BHs is still a matter of debate (e.g., Sahu et al. 2022), it is worth exploring its possible impact in the predicted scaling relations in a full and self-consistent galaxy and supermassive BH co-evolution model (see, e.g., Barausse et al. 2017).
 
The paper is organized as follows. In Sect. 2 we first summarize the basic features of the semi-analytic galaxy formation model. The implementation of the above two-dimensional description of AGN outflows in the semi-analytic model is presented in Sect. 2.2, while in Sect. 2.3 we describe how we compute the bulge stellar velocity dispersion for each model galaxy. In Sect. 3 we briefly recall how the computed stellar velocity dispersion allows to study angular resolution-related selection effect affecting the comparison between the observed and the predicted scaling relation between the dynamically-measured BH mass and the galaxy properties. In Sect. 4 we present our results for 
the local scaling relations between the BH mass and both the bulge and the global properties of the host galaxies, and we present our predictions for the redshift evolution of the $M_{BH}-M_*$ relation. 
Sect. 5 is  devoted to discussion and conclusions. 
  
\section{Method} 
We base our present work on the semi-analytic model (SAM) described in previous papers (see Menci et al. 2005; 2014, 2016). However, the specific version of the SAM adopted here differs from the one presented in the above papers since it implements a new, detailed description of AGN feedback, as we shall discuss in detail below (Sect. 2.1), and is extended to describe the galaxy stellar velocity dispersion (Sect. 2.2). 

The backbone of the model is based on the merging trees of the DM halos, which are generated through a Monte Carlo procedure, with merging probabilities 
given by the extended  Press \& Schechter formalism (see Bond et al. 1991; Lacey \& Cole 1993), assuming a CDM power spectrum of perturbations in a 
concordance cosmology with density parameter $\Omega_0=0.3$, a baryon density parameter $\Omega_b=0.04$, a dark energy density parameter $\Omega_\Lambda=0.7$, and a Hubble constant $h$=0.7

 The sub-halos included into the main halo can coalesce 
with the central galaxy after the orbital decay due to dynamical friction, or
merge with other satellite sub-halos, as described in our previous papers (Menci et al. 2005, 2014, 2016). Gas cools inside the halos due to radiative processes, and settles into a rotationally supported disc with mass $M_{gas}$, with scale length $r_d$ and circular velocity $v_d$ related to the DM circular velocity $V_c$ as given in Mo, Mao, White (1998). The stars are converted from the gas through three channels: (1) quiescent star formation with long time scales: $\sim1$ Gyr; (2) starbursts following galaxy interactions with timescales $\lesssim 100$ Myr, according to BH feeding; (3) the loss of angular momentum triggered by the internal disk instabilities causing the gas inflows to the center, resulting in stimulating star formation (as well as BH accretion). The stellar feedback is also considered by calculating the energy released by the supernovae associated with the total star formation which returns a fraction of the disc gas into a hot phase. A Salpeter IMF is adopted in the SAM simulation.

We assume BH seed $M_{seed}=100\,M_{\odot}$ (Madau \& Rees 2001)  to be initially present in all galaxy progenitors at the initial redshift $z=15$. This constitute an approximate way of rendering the effect of the collapse of PopIII stars. However, the detailed value of $M_{seed}$ has a negligible impact on the final BH masses as long as they remain in the range $M_{seed}=50-500\,M_{\odot}$.

The BH accretion is based on interactions feeding mode and disk instabilities feeding mode.
\newline
(1) {\it triggered by interactions.}  Each model galaxy (with tidal radius $r_t$ and circular velocity $V_c$) in our Monte Carlo simulation may interact inside a host DM halo with circular velocity $V$ at a rate $\tau_r^{-1}=n_T\,\Sigma (r_t,V_c,V)\,V_{rel} (V)$. Here $V_{rel}$ is the average relative velocity of sub-halos inside a host DM halo wioth circular velocity $V$, and $n_T$ is their number density in the common DM halo. The interaction rate  determines the probability for encounters, either fly-by or merging, through the corresponding cross sections $\Sigma$ given in Menci et al. (2014).  The fraction $f=(1/2)|\Delta j/j|$ of gas destabilized in each interaction corresponds to the relative loss $\Delta j$ of orbital angular momentum $j$. In the above equation, the pre-factor accounts for the probability 1/2 of inflow rather than outflow related to the sign of $\Delta j$. Both the destabilized gas fraction $f$ and the cross section $\Sigma$ depend on DM the mass of the considered galaxy $M_h$, and on the DM mass $M_h'$ of the partner galaxy in the interaction, extracted in our Monte Carlo procedure. Both major mergers ($M_h \sim M_h'$) and  minor mergers ($M_h \gg M_h$')are considered.
\newline
(2) {\it induced by disc instabilities.} We assume these to arise  in  galaxies with disc mass exceeding $M_{crit} =  {v_{max}^2 r_{d}/ G \epsilon}$ with $\epsilon=0.75$, where $v_{max}$ is the maximum circular velocity associated to each halo (Efstathiou, Lake, Negroponte 1982). Here $\epsilon \approx 0.5-0.75$ is a parameter calibrated on simulations. The above criterium is that usually adopted in semi-analytic models: Hirschmann et al. (2012) adopt a value $\epsilon = 0.75$, and a similar value is adopted by Fanidakis et al. (2011). In order to investigate the maximal affect of DI on the statistical evolution of AGN we also adopt the value $\epsilon = 0.75$.
Such a criterion strongly suppresses the probability for disc instabilities to occur not only in massive, gas-poor galaxies, but also in dwarf galaxies characterized by small values of the gas-to-DM mass ratios. The instabilities induce loss of angular momentum resulting into strong inflows that we compute following the description in Hopkins (2011), recast and extended as in Menci (2014). 

The BH accretion rates associated to the  two feeding modes above yields and 
AGN emission  with bolometric luminosity $L_{AGN}=\eta\,\dot M_{BH}\,c^{2}$, where we assume the accretion efficiency to take the value $\eta=0.1$
(Yu \& Tremaine 2002; Marconi et al. 2004; Shankar et al. 2020a,b). The radiation field of the AGN 
pushes the cold gas disc outwards, determining the expansion of a shock front  
 (Silk \& Rees 1998; Cavaliere, Lapi, Menci 2002; King 2003; Lapi, Cavaliere, Menci 2005; Granato et al. 2004; Silk \& Nusser 2010; King, Zubovas \& Power 2011; Zubovas \& King 2012;  Faucher-Giguere \& Quataert 2012; King \& Pounds 2015). This results in the expulsion of a fraction of disc gas (AGN feedback) which is also described in the model. However, we have substantially upgraded such a description as shown in detail in the next subsection below. 

The above model has been upgraded in two fundamental respects: \newline
1) the description of the AGN feedback is now performed on the basis of a complete two-dimensional model that describes the expansion of the blast wave associated to 
the outflow in a disc geometry, and the outflow mass is computed from such a 
model integrating the mass outflow rate over the directions where the outflows escapes from the galactic disc; \newline
2) we have implemented a description of the bulge size and of the velocity dispersion of stars in the bulge. 

We describe the two upgrades in turn. 

\subsection{Implementing a two-dimensional physical description of outflows in galactic discs}

In a previous paper (Menci et al. 2019) we provided a compact two-dimensional description for the expansion of AGN-driven shocks in realistic galactic discs with exponential gas density profiles in a disc geometry.  The gas distribution is assumed to be axisymmetric (with the $X$-axis aligned with the plane of the disc and the $Y$ coordinate aligned with the rotation axis), with a density profile $\rho(r)=\rho_0\,exp(-r/r_d)$ depending only on the galacto-centric distance $r$ and on a scale length $r_d$, but with a cutoff in the $Y$ direction at a distance corresponding to the disc scale-height $h$, i.e., $\rho=0$ for $Y\geq h$.  The normalization $\rho_0$ is such as to obtain the 
disc gas mass $M_{gas}$ when integrated out to large radii (i.e. $r\rightarrow \infty$), 
while the scale-height is assumed to be constant with radius for a given galaxy, and to increase with the galaxy circular velocity according to the observed average relation 
$h=0.45\,(V_c/100 {\rm km/s} - 0.14$ kpc (see van der Kruit \& Freeman 2011 and references therein). Details are given in Menci et al. (2019). 

The description accounts for the balance between the pressure term (fueling the expansion) acting on the surface element corresponding to the solid angle in the considered direction, and the counter- acting gravitational term, determined by the total mass $M$ (contributed by the DM and by the central BH) within the shock radius. It includes the cooling of both the shell and gas inside the bubble due to the relevant atomic processes (inverse Compton and free-free emission, see Richings, Faucher-Giguere 2018). 

We derived solutions to the outflow velocity $V_{S,\theta}$, mass outflow rate $\dot M_{S,\theta}$ and shock position $R_{S,\theta}$ as a function of time in different directions defined by the angle $\theta$ with respect to the plane of the disc. These depend on three fundamental quantities related to host galaxy: 
the AGN bolometric luminosity $L_{AGN}$ (fueling the expansion),  
the mass of cold gas in the disc $M_{gas}$ (the mass that is pushed by the 
wind), the total circular velocity $V_c$ (determining the depth of the potential wells that counter-act the expansion). 

\vspace{0.2cm}
\hspace{-0.4cm}
\scalebox{0.3}[0.3]{\rotatebox{0}{\includegraphics{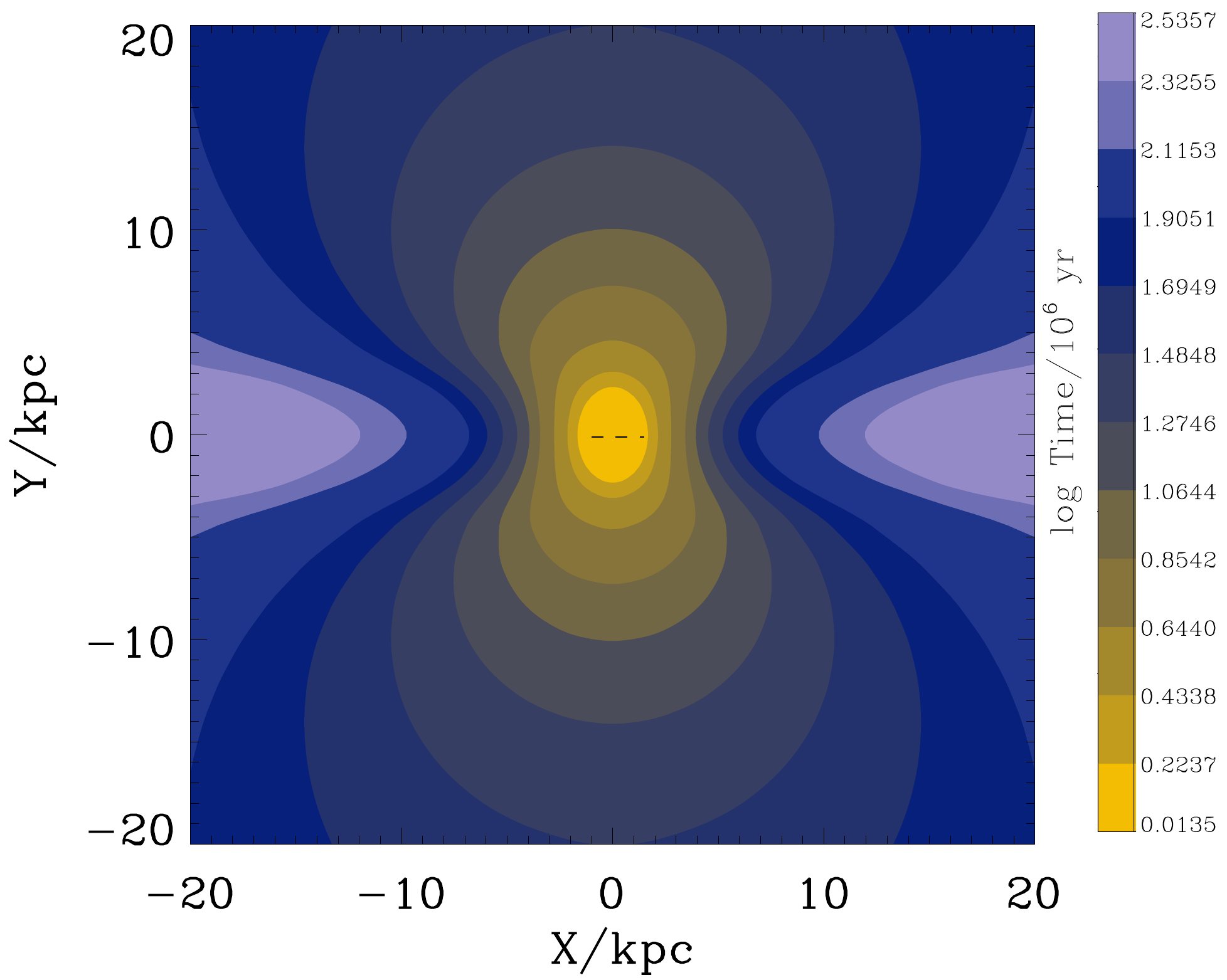}}}
\scalebox{0.3}[0.3]{\rotatebox{0}{\includegraphics{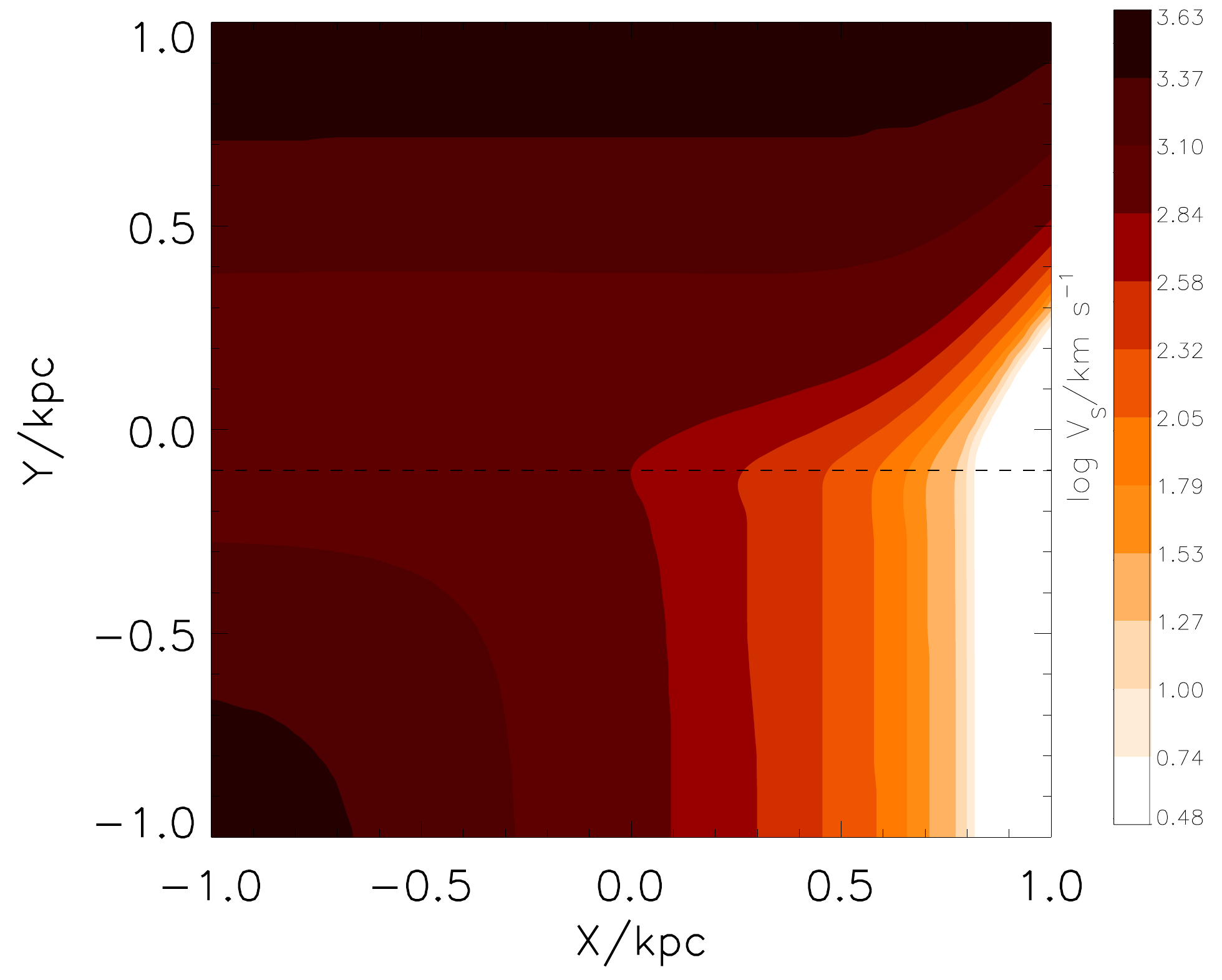}}}
\scalebox{0.3}[0.3]{\rotatebox{0}{\includegraphics{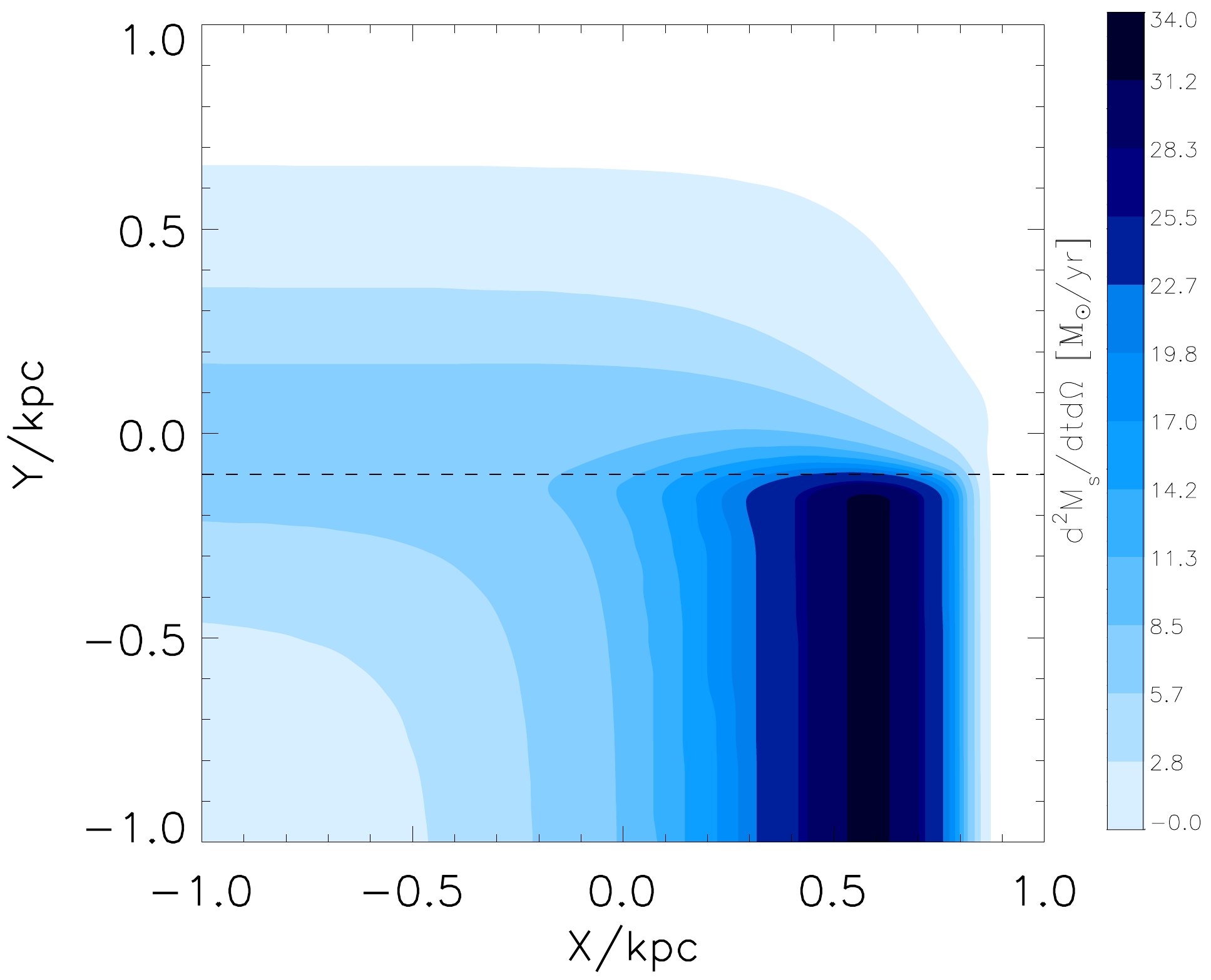}}}
\vspace{0.2cm }\newline
{\footnotesize Fig. 1. Left Panel: for our reference galaxy, we show the positions of the shock at the different times represented by different colors and displayed on the right bar. The central and the 
right panel show the  velocity map and mass outflow rate map, respectively. The values corresponding to the colored contours are displayed on the bars. The $X$ and $Y$ coordinates correspond to the distance from the galaxy center in the directions parallel and perpendicular to the plane of the disc, respectively.  }
\vspace{0.2cm }

The solutions  are plotted in Fig. 1 for a reference galaxy with DM mass $M_h=10^{12}\,M_{\odot}$, a gas 
mass  $M_{gas}=10^{10}\,M_{\odot}$, and AGN bolometric luminosity $L=10^{45}$ erg/s. The $X$ coordinate represents the distance from the center in the direction parallel to the plane of the disc, while the $Y$ coordinate corresponds to the distance in the (vertical) direction perpendicular to the disc. 

The shock expansion radius follows the paths of  least resistance (see left panel of Fig. 1), yielding an elongated shock front in the vertical direction. Inspection of Fig. 1 shows that while in the direction perpendicular to the disc the outflows reaches a distance of 20 kpc in approximatively $10^{7}$ yrs, it takes about $10^{8}$ yrs to reach the same distance in the plane of the disc.
This has important implications for studies of AGN feedback in galaxy formation models. E.g., for an AGN 
 life time$\sim 10^{8}$ yrs this would results into null gas expulsion along the plane of the disc. 

 Along the plane of the disc the velocity $V_{S,\theta}$ rapidly decreases with increasing radius (central panel), while in the vertical direction the shock decelerates until it reaches the disc vertical boundary $h$, but it rapidly accelerates afterward due to the drop of the gas density outside the disc. The opposite is true for the mass outflow rate (right panel), which instead grows appreciably only along  the plane of the disc, where the larger densities allow to reach values $\dot M_{S,\theta=0}\sim 10^{3}$ $M_{\odot}$/yr. 

The above shock quantities  depend on the global properties of the host galaxy and of the central AGN.  For a given set of galaxy properties 
(cold gas mass $M_{gas}$ in the disc, circular velocity of the DM halo $V_c$ and AGN luminosity $L_{AGN}$) the predictions of the model (mass outflow rates and outflow velocities in any direction) can be compared with observations. In Menci et al. (2019) we have tested the model  against a state-of-the-art compilation (Fiore et al. 2019) of observed outflows in 19 galaxies with different measured gas and dynamical mass, allowing for a detailed, one-by-one comparison with the model predictions. 
Specifically, for each observed object, the measured AGN and host galaxy properties (AGN luminosity, gas mass and circular velocity) have been used to obtain the input quantities for the model. The resulting velocity and mass outflow rates computed from the model have then been compared with the observed corresponding wind properties. The model yields - for each considered galaxy and at the observed outflow radii - values of velocity and mass outflow rates (averaged over the directions) that are in good agreement with observations.  The predicted densities of the shocked shell are consistent with the observed molecular emission of the outflows in the vast majority of cases.
The agreement we obtained - for a wide range of host galaxy gas mass ($10^{9}\,M_{\odot}\lesssim M_{gas}\lesssim 10^{12}\,M_{\odot}$) and AGN bolometric luminosity
 ($10^{43}\,{\rm erg\,s^{-1}}\lesssim L_{AGN}\lesssim 10^{47}{\rm erg\,s^{-1}}$) - provides a  quantitative systematic test  for the modeling of  AGN-driven outflows in galactic discs. It  provides a solid baseline for a reliable  implementation of AGN feedback in galaxy formation models. 
 
Implementing the above modeling of AGN outflows in the SAM constitutes a challenging task. Directly solving the equation for the expansion of outflows in each of the simulated galaxies generated by our SAM would be too demanding in terms of computation time. Thus, we proceed as follows.  

We consider a grid of values for the three input quantities of the outflow model.  
Specifically, we consider 20 equally spaced logarithmic values for the AGN luminosity in the range $log\,(L_{AGN}/{\rm erg\,s^{-1}})=42-47$, for the gas mass  in the range $log \, (M_{gas}/M_{\odot})=8-11.5$, and for the circular velocity of the host galaxy $log\,(V_c/{\rm 100 km/s})=0.5-4$. 

For each combination ($L_{AGN}$, $M_{gas}$, $V_c$) we run our model for the expansion of the outflows and compute  $R_{S,\theta}$, 
$V_{S,\theta}$, and mass outflow rate $\dot M_{S,\theta}$ as a function of time in different directions defined by the angle $\theta$. We then compute the total 
outflow mass $M_{S}$ integrating $\dot M_{S,\theta}$ over time and over  all the directions where the outflow escapes, i.e., the directions where velocity $V_{S,\theta}$ accelerates after reaching the disc boundary (see Fig. 1) to reach velocities well beyond the escape velocity from the galactic halo. 
This provides us with tabulated values of $M_{S}$ (and of the critical opening angle $\theta_{c}$ corresponding to escaping outflows) for each combination 
 ($L_{AGN}$, $M_{gas}$, $V_c$).  The expelled gas fraction $f_{expell}=M_S/M_{gas}$ and the opening angle $\theta_c$ as a function of $L_{AGN}$ and $M_{gas}$ are shown in Fig. 2 for a reference value of the circular velocity $V_c=200$ km/s.
\vspace{0.2cm}
\begin{center}
\scalebox{0.33}[0.33]{\rotatebox{0}{\includegraphics{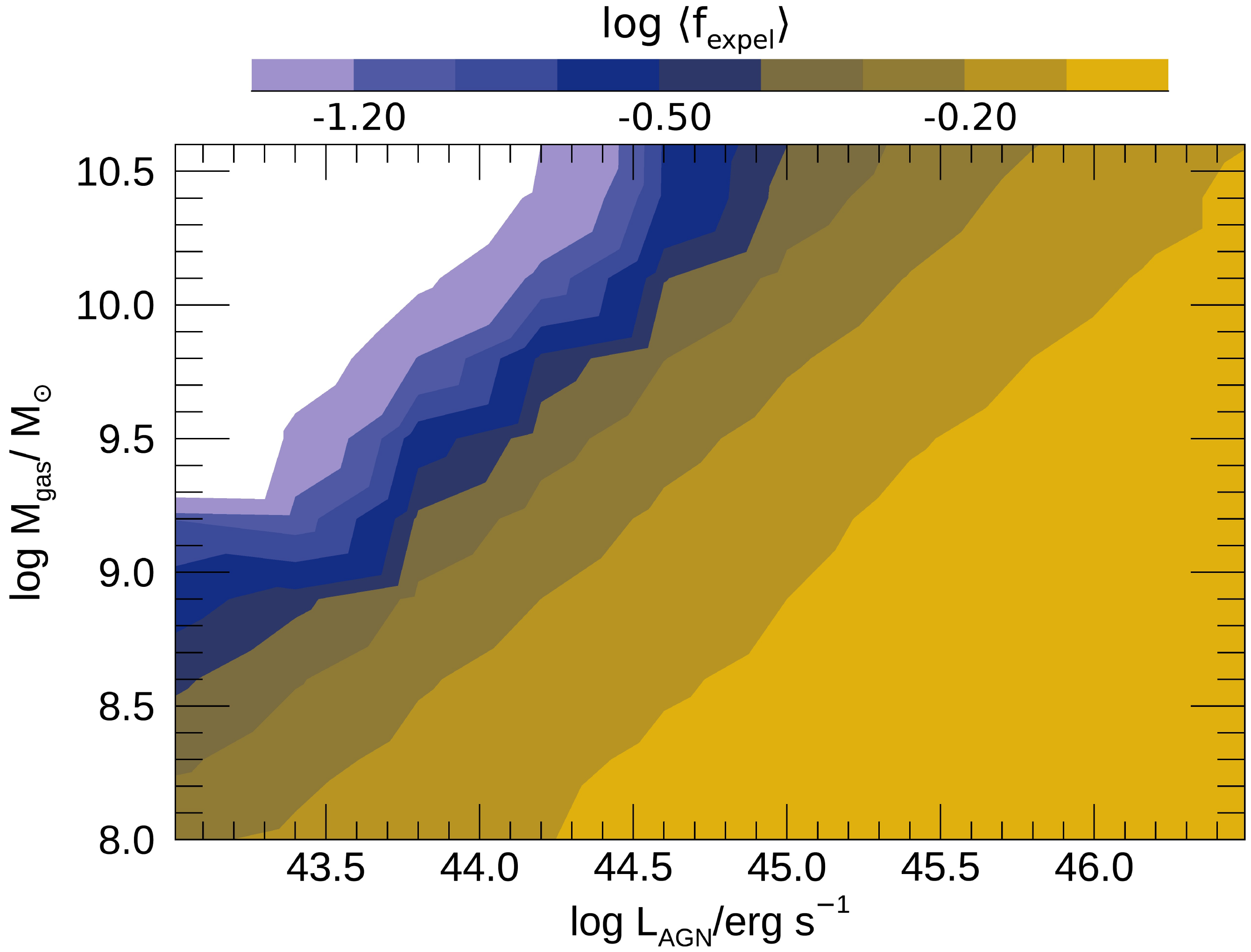}}}
\scalebox{0.33}[0.33]{\rotatebox{0}{\includegraphics{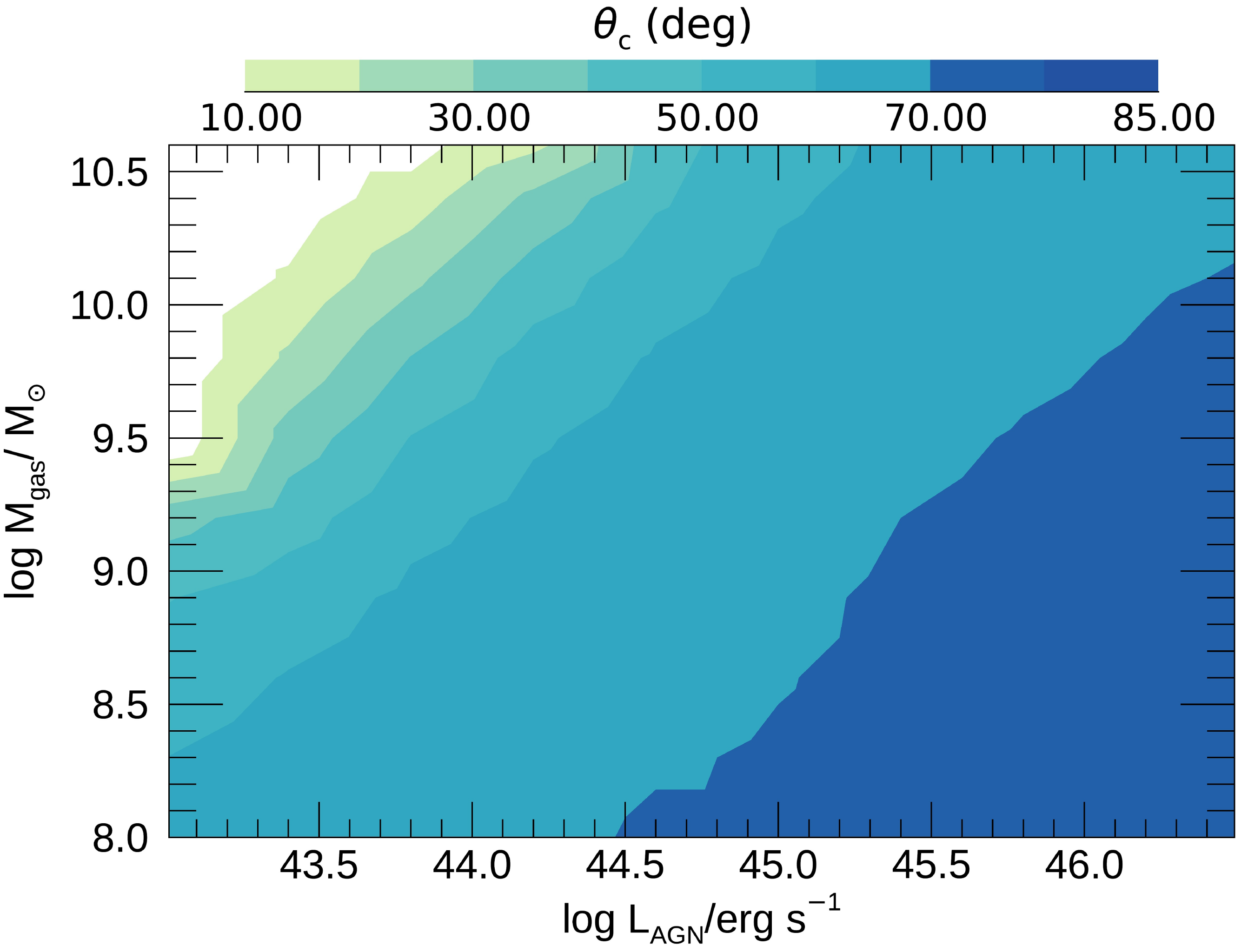}}}
\end{center}
\vspace{0.2cm }
{\footnotesize Fig. 2.  Left Panel. The  color code shows the dependence of the fraction of cold gas expelled by AGN outflows ($f_{expell}$, left panel) and of the solid angle $\theta_c$ swept out by AGN outflows (right panel) on the AGN bolometric luminosity $L_{AGN}$ and of the cold gas mass $M_{cold}$.  A DM circular velocity  $V_c=200$ km/s is assumed for both panels.}
\vspace{0.2cm }

To compute the effect of AGN feedback on each galaxy of our SAM, we interpolate the above tabulated values of $M_{S}$ and $\theta_c$ to find the escaped gas mass and opening angle corresponding to the  combination ($L_{AGN}$, $M_{gas}$, $V_c$)  associated to each model galaxy in the SAM. The distribution of the average  values of the expelled gas fraction $f_{expell}$ as a function of the stellar mass $M_*$ and star formation rate $\dot M_*$ is shown in Fig. 3. 

The distribution shows a bimodal shape at low values of $M_*$. In fact, large values of $f_{expell}$ require either low values of $M_{gas}$ (and hence low values of $\dot M_*$), and/or large values of $L_{AGN}$ (attained for large gas mass $M_{gas}$, yielding on average large values of $\dot M_*$). For large values of  $M_*$, the outflows can overcome the effect of the deep galaxy potential wells 
(yielding large expelled fractions $f_{expell}\approx 1$) only when powered by a large AGN energy injection $L_{AGN}$, which can only be attained for large values of $M_{gas}$ and hence of $\dot M_*$. The inclusion of a kinetically-driven AGN feedback, can thus trigger powerful outflows in massive galaxies, that can push large portions of the gas in the intergalactic medium. This expelled gas will not be re-accreted at later times. We have checked that such a feedback model yields stellar galaxy luminosity functions consistent with observations, although in this respect the isotropic feedback model previously adopted in our SAM performed equally well.  
 
\vspace{0.cm}
\begin{center}
\scalebox{0.48}[0.48]{\rotatebox{0}{\includegraphics{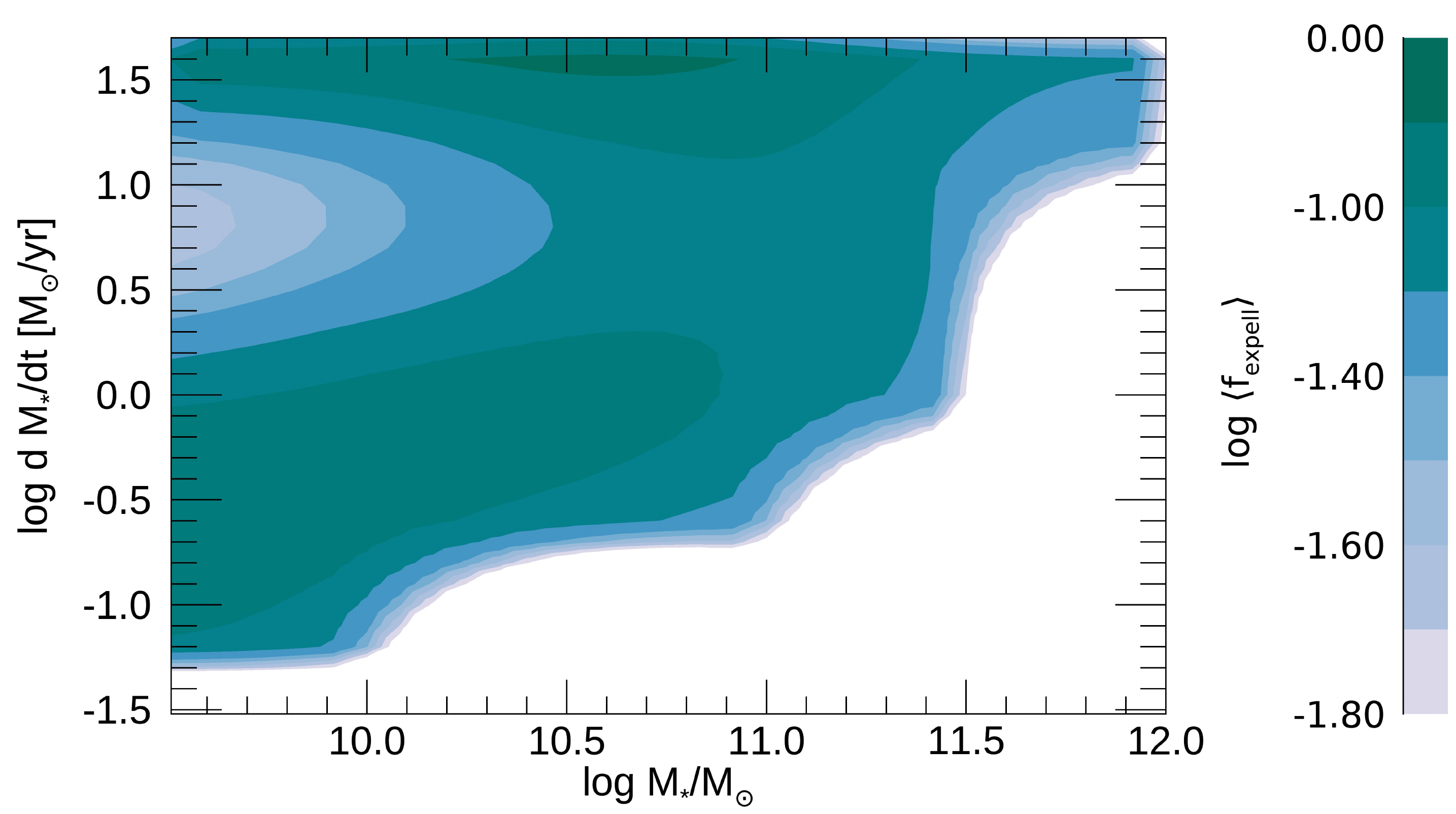}}}
\end{center}
\vspace{0.1cm }
{\footnotesize Fig. 3. The color code shows the average fraction of gas expelled by AGN outflows  in galaxies with different stellar mass $M_{*}$ and star formation rate $\dot M_*$. }
\vspace{0.1cm }

\vspace{-0.4cm}
\hspace{-0.8cm}
\begin{center}
\scalebox{0.6}[0.6]{\rotatebox{0}
{\includegraphics{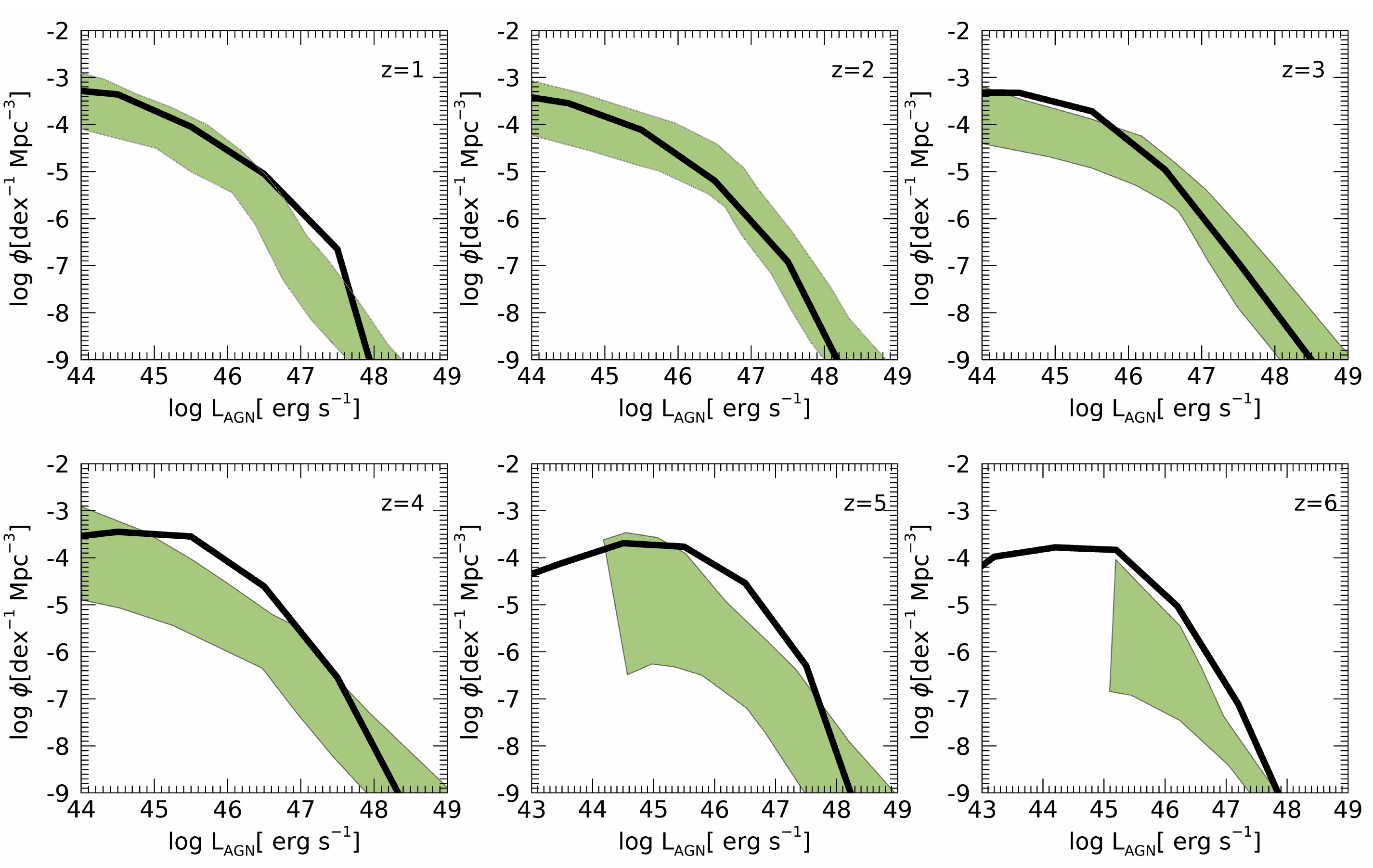}}}
\end{center}
\vspace{0.1cm }
{\footnotesize Fig. 4.  The predicted  AGN bolometric luminosity functions (solid lines) are compared to 
the compilation of data in Shen et al. (2020, shaded regions) in six redshifts bins shown in the labels. 
}
\vspace{0.3cm }

We have checked that the new treatment of AGN feedback described above yields AGN luminosity functions 
consistent with existing observations. This is shown in Fig. 4, where the predicted AGN bolometric luminosity function is compared with the compilation by Shen et al. (2020) in different redshift bins extending up to $z=6$. The slight over-prediction occurring at $z\gtrsim 3$ is not surprising. In fact, 
 the predicted luminosity functions include Compton-thick AGN, which  constitute a relevant fraction of the AGN population, with an appreciable uncertainty range $\sim 20-50$ \% at $z\gtrsim 3$ (Shen et al. 2020) which is consistent with the slight over-estimate of the observed abundance in the same redshift range. In addition, large observational uncertainties still affect the measurement of the faint end of the AGN luminosity function at $z\gtrsim 3$ (see, e.g., Shankar \& Mathur 2007; Ricci et al. 2017; Shen et al. 2020). Based on results by recent surveys, the number of quasars at redshifts $z \gtrsim 3$ is being constantly revised upward, with several authors  obtaining abundances of low-luminosity AGN up to 40 \% larger than the estimates reported in the analysis by Shen et al. (2020) at both faint ( $M_{UV}\approx -23.5$, Boutsia et al. 2018; Giallongo et al. 2019) and bright ($M_{UV}\approx -27$, Boutsia et al. 2021;  Grazian et al. 2020) magnitudes. 
 
Having tested the  new treatment of AGN feedback in the SAM yields AGN luminosity functions consistent with existing observations over a wide range of redshift, we proceed with the extension of the SAM to include a description of the stellar  velocity dispersion, as detailed below. 

\subsection{Computing Jeans-based stellar velocity dispersions}

In this work, we have further extended our SAM with respect to previous renditions of the model to include state-of-the-art calculations of the half-mass radii and stellar velocity dispersions. The former have been presented in Zanisi et al. (2020). Half-mass galactic radii in progenitor discs are computed from angular momentum conservation with the host halo spin. When galaxies merge, a stellar bulge  is formed with the half-mass radius $R_h$,  which is computed via energy conservation between the gravitational potential energies of the progenitors and the descendant (e.g., Cole et al. 2000).
We have tested that the sizes computed in our SAM match the observed correlation with stellar mass for the redshift range covered by observations, for both the star-forming and the quiescent galaxies.  
To perform the comparison, we adopt the approach in  Shankar et al (2014). 
Assuming that light traces mass, the above authors 
 convert $R_{h}$ to projected  half-light radii $R_e$ using the tabulated factors from Prugniel \& Simien (1997), i.e., $R_{e} \approx 2S(n)\,R_{h}$, with the scaling factors $S(n)$ dependent on the Sersic index $n$. Following 
 Shankar et al (2014), for the latter we adopt a fixed value $n=4$. As noted by the above authors, while in principle   it is possible to predict a Sersic index a priori from the models (e.g., Hopkins et al. 2009),  this relies on several additional assumptions on the exact profile and its evolution with time of the dissipational and dissipationless components, that the true advantage with respect to simply empirically assign a constant Sersic index is modest. By the same token, we note that none of our main results would change if we were to assign to our mock galaxies stellar mass-dependent Sersic indices and/or effective radii directly extracted from the observed distributions.

\vspace{-0.4cm}
\begin{center}
\scalebox{0.55}[0.55]{\rotatebox{0}{\includegraphics{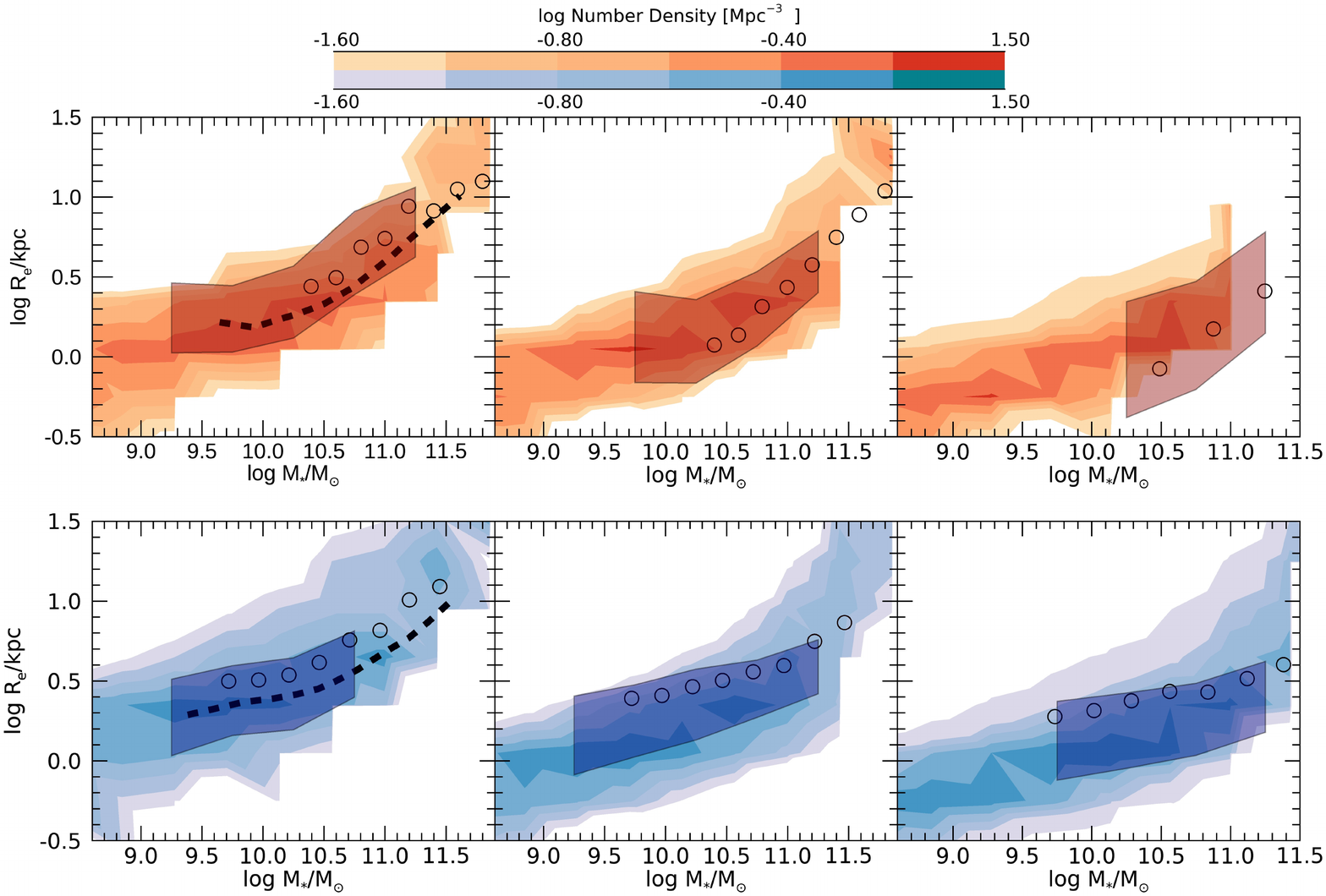}}}
\end{center}
\vspace{-0.4cm }
{\footnotesize Fig. 5.  The predicted distribution of  effective radius $R_e$ shown as a function of the stellar mass $M_*$ at redshifts $z=0.25$, $z=1.5$1, and $z=2.25$ (from left to right),  for early -type  (top row) and late-type galaxies (bottom row). 
The predictions are compared with the data from Van der Wel 2014 (shaded areas). We also show as dashed lines and as dots the size-stellar mass relation measured at $z=0.1$  Bernardi et al. (2014), and by Mowla et al. (2019), respectively }
\vspace{0.2cm }

The predicted correlation between the size $R_e$ and the stellar mass is compared with observations in Fig. 5 for different redshift bins, and for both quiescent and star-forming galaxies. 
Following the approach in Genel et al. (2018), we classify our SAM-generated galaxies as early-type or late-type through their specific star formation $SSFR=\dot M_*/M_*$. Galaxies are defined as ``quenched'' if their SSFR is at least 1 dex below the ridge of the star formation main sequence. This is 
defined, for simplicity, as $log (SSFR/Gyr^{-1})$=-0.94, -0.85, -0.35, 0.05 for $z$=0, 0.1, 1, 2, respectively. On the observational side, 
van der Wel et al. (2014) classify early-type and late-type galaxies on the basis of their star-formation properties, 
defined in terms of cuts in the (U-V)-(V-J) rest-frame colors, while Bernardi et al. (2014) adopt a morphological 
classification based on a  Bayesian automated procedure. 
Notice that we assume that our model half-light radius $R_e$ is computed assuming that light traces mass. On the other hand, the different observations are made in different bands. While  Bernardi et al. (2014) measurements are performed  directly in the Sloan Digital Sky  Survey (SDSS) r band, the ones from van der Wel et al. (2014) are made in the Hubble Space Telescope Wide Field Camera 3 (WFC3) F125W filter, centred on an observed- frame wavelength of 1.25$\mu$m. In order to match the SDSS-based studies, we convert the measurements in  van der Wel et al. (2014) to the SDSS r band using the gradient $d log R_e/d log \lambda$  they directly measure between the F814W (814 nm) and F125W filters.

From the galaxy size computed as shown above, we can derive the  corresponding velocity dispersion solving the spherical Jeans equation, following Desmond \& Wechsler (2017). For the bulge component, this reads 
\begin{equation}
{d\rho (r)\,\sigma^{2}_{r}\over dr}+{2\,\beta(r)\over r}\rho(r)\,\sigma^{2}_{r}=-G\rho(r)\,{M_{*,b}(r)\over r^{2}}
\end{equation}
where $\beta(r)=1-\sigma^{2}_{\theta}/\sigma^{2}_r$ is the orbit anisotropy profile relating the radial velocity dispersion  $\sigma_r$ to the tangential component  $\sigma_{\theta}$,  $\rho(r)$ is the mass density profile of the bulge, and $M_{*,b}$ is the bulge stellar mass. 

The line-of-sight velocity dispersion solving  eq. 1 is then given by Mamon \& Lokas (2005; see also Desmond \& Wechsler 2017) in the form 
\begin{equation}
\sigma_{los}^{2}(R)={2\,G\over \Sigma (R)}\,\int_{R}^{\infty}\,K(r/R)\,\rho(r),
M_{*,b}(r)\,{dr\over r}
\end{equation}
where $R$ is the projected distance from the center,  $\Sigma (R) $ is the projected density profile. The function $K$ is defined as a combination of 
gamma functions $\Gamma$ and incomplete beta functions $B$:  
\begin{equation}
K(u)\equiv 1/2\,u^{2\beta-1}\,[(3/2-\beta)\sqrt{\pi}\,(\Gamma(\beta-1/2)/\Gamma(\beta))+\beta B(1/u^{2},\beta+1/2,1/2)-B(1/u^{2},\beta-1/2,1/2)]
\end{equation}
where $u\equiv r/R$, and $\beta$ is the assumed orbit anisotropy. 

For each model galaxy, we assume for $\Sigma (R) $ a Sersic form $\Sigma (R)=\Sigma_0\,g(R)$ with  
\begin{equation}
g(R/R_{e})=e^{
{-b_n\,
\big[
\big(
{R\over R_{e}}
\big)^{1/n}-1
\big]}
}, 
\end{equation}
where $b_n\approx 2n-1/3+0.009876/n$. 
This is normalized as to yield the bulge mass $M_{*,b}$ when integrated over $R$, so that $\Sigma_0=(1/2\pi\,Q)\,M_{*,b}/R_{e}^{2}$ where 
$Q\equiv \int_{0}^{\infty}g(y)\,y\,dy$. 

The corresponding  de-projected mass density profile is $\rho (r)=\rho_0\,f(r/R_{e})$ 
where 
\begin{equation}
 f(r/R_{e})=
\Bigg({r\over R_{e}}\Bigg)^{-p_n}
e^{
{-b_n\,
\big(
{r\over R_{e}}
\big)^{1/n}
}
}
\end{equation} 
where $p_n=1-0.6097/n+0.00563/n^2$. Again, the normalization $\rho_0$ is computed so as to obtain the total  mass $M_{*,b}$ when the above mass density profile is integrated over the volume. Thus, 
$\rho_0=(1/4\pi\,F)\,M_{*,b}/R_{e}^{3}$ where 
$F\equiv \int_{0}^{\infty}g(y)\,y^2\,dy$. 

Finally, the bulge mass density profile is $M_{*,b}(r)=4\,\pi\,\int_{0}^{r}\,\rho(y)\,y^{2}\,dy$. Inserting the above form for the density profile $\rho(r)$ 
we obtain 
\begin{equation}
M_{*,b}(r/R_e)=
(M_{*,b}/F)\,\int_{0}^{r/R_{e}}\,y^{2}f(y)\,dy\equiv (M_{*,b}/F)\,h(r/R_{e})
\end{equation}

Inserting eqs. (4), (5), (6) in eq. 2 we obtain  
\begin{equation}
\sigma_{los}^{2}(x)={G\,M_{*,b}\over R_{e}}\,{1\over F\,Q} \, 
\int_{x}^{\infty}\,K(y/x)\,f(y)\,h(y)\,dy/y
\end{equation}
where $x\equiv R/R_{e}$. 

In the following, we shall consider the average velocity dispersion 

\begin{equation}
\sigma^{2}={\int_{0}^{\infty}\Sigma(x)\sigma^{2}_{los}x\,dx\over 
\int_{0}^{\infty}\Sigma(x)\,x\,dx}
\end{equation}
Notice that when a finite aperture radius $R_{ap}$ is considered, 
the upper limits in the above integrals should be replaced with $x_{ap}\equiv R_{ap}/R_e$. Inserting the expression in eq. (7) we obtain
\begin{equation}
\sigma^{2}= {G\,M_{*,b}\over R_{e}}\, W
\end{equation}
with a 'shape-factor'  $W$  defined as 

\begin{equation}
W(\beta,n) \equiv{1\over F\,Q} \, {\int_{0}^{\infty}\,g(x)\,x\,dx
\int_{x}^{\infty}\,K(y/x)\,f(y)\,h(y)\,dy/y\over 
\int_{0}^{\infty}\,g(x)\,x\,dx}
\end{equation}

The velocity dispersion is thus related to the virial 
value $GM_{*,b}/R_{e}$ through the 'shape-factor' 
$W(\beta,n)$ which depends on the structural properties of the bulge 
(anisotropy parameter $\beta$ and Sersic index $n$). 

For each simulated galaxy in our SAM, the effective bulge mass content $M_{*,b}$ and radius $R_{e}$ allow us to derive the average bulge velocity dispersion from the relations (9), (10), for a given choice of 
$\beta$ and $n$. For the first, we extract a random value 
from a Gaussian distribution with average  value $\langle \beta\rangle =0.3$ and dispersion 0.35. Such a choice has been shown to provide a good fit to observational data  (Desmond \& Wechsler 2017). As for the Sersic index $n$, 
we adopt a fixed vale $n=4$ based on the considerations presented above.  
To  test that the treatment above is consistent with available data, we have compared (Fig. 6) the velocity dispersion of passive model galaxies ($\dot M_*/M_*\leq 0.1$ Gyr$^{-1}$) at low redhsift ($z\leq 0.1$) to existing data for early-type galaxies in the SDSS data base (Bernardi et al. 2011), selected  on the basis of morphological indicators  (Hyde \& Bernardi 2009). The observational correlation does not change appreciably if passive galaxies are selected on the basis of spectral or color indicators (Belli et al. 2014; Zahid et al. 2016). Such galaxies are characterized by large bulge-to-disc stellar mass ratios $B/T\gtrsim 0.8$, for which the total galaxy velocity dispersion is dominated by the bulge component computed in eq. 8-10. For consistency with the data by Hyde \& Bernardi (2009), when computing eq. 8 we have adopted $x_{ap}=1/8$ (this leads to a small offset -0.03 in $\log (\sigma)$ when compared to the global average corresponding to $x_{ap}\rightarrow \infty$). 
The excellent agreement shown in  Fig. 6 provides a reliable basis for computing the scaling relations 
 between the BHs and the properties of the host galaxies, as well as for including the effects of observational selection biases in the analysis of model results, as we discuss below. 

\vspace{-0.1cm}
\begin{center}
\scalebox{0.53}[0.53]{\rotatebox{0}{\includegraphics{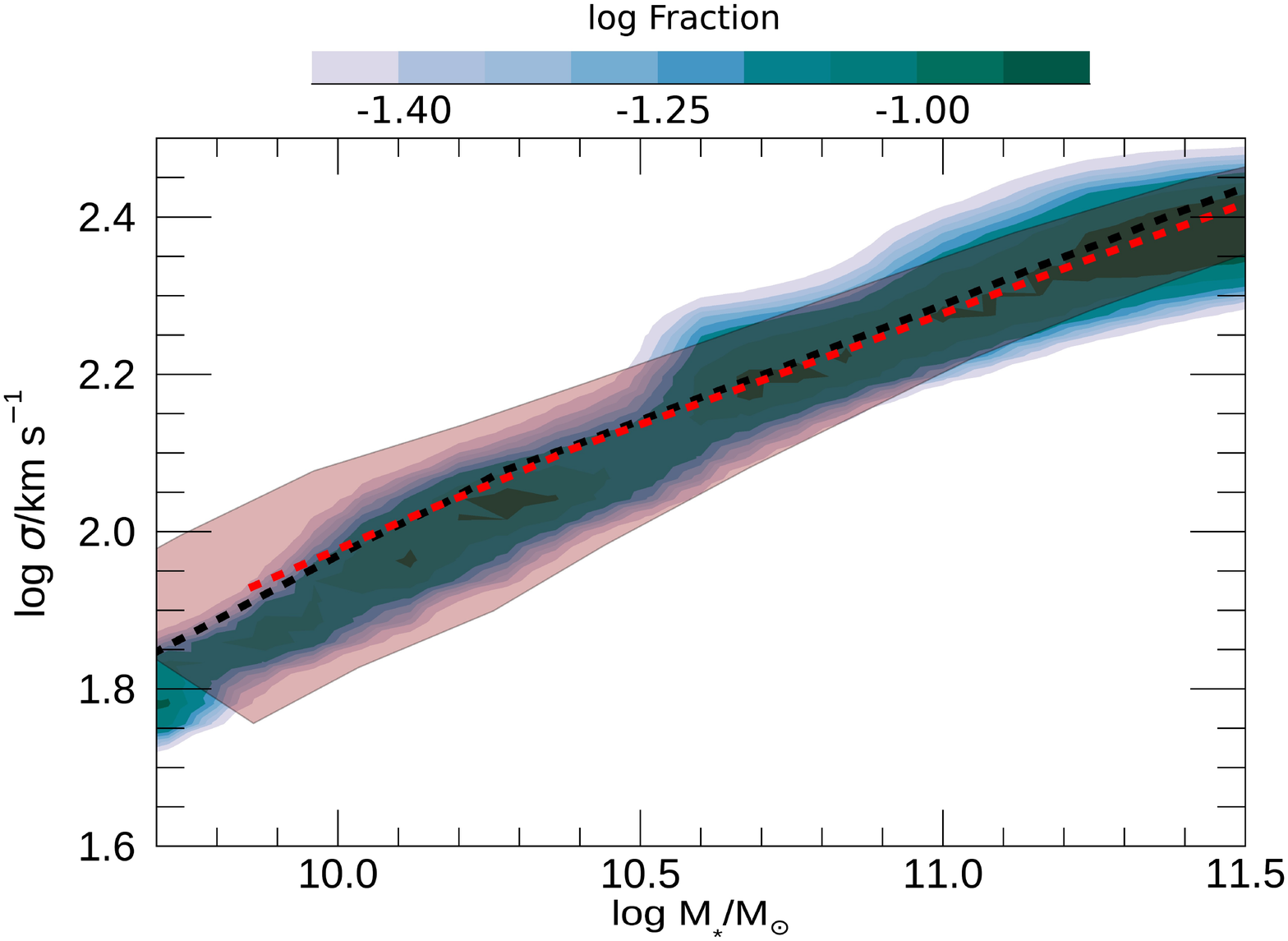}}}
\end{center}
\vspace{-0.4cm }
{\footnotesize Fig. 6.  The predicted relation between the velocity dispersion $\sigma$ and the stellar mass $M_*$ for passive galaxies (see text).  
The color code represents the logarithm of the fraction of model galaxies with the considered $\sigma$ for each bin of $M_*$. The shaded band shows the relation for early-type galaxies selected on the basis of morphological indicators after Hyde \& Bernardi (2009). 
 The dashed black and red lines is the average relation derived by Zahid et al. (2016) for passive SDSS galaxies and for the {\it Smithsonian Hectospec Lensing Survey} (SHELS) sample,  selected on the basis of spectral properties ($D_n4000>1.5$, where $D_n4000$ is  the flux ratio between two spectral windows adjacent to the $4000 {\AA}$ break, see Balogh et al. 1999). }
\vspace{0.2cm }

\section{Selection Bias} 

Shankar et al. (2016) have shown that a selection bias may affect the 
observe scaling relations, when dynamical measurements of the BH mass are considered. In fact, such mass estimates are only possible if the black hole’s sphere of influence 
\begin{equation}
r_{infl} \equiv G\,M_{bh}/\sigma^{2}
\end{equation}
has been resolved in the observations (e.g. Peebles 1972;  Gultekin et al. 2011; Graham \& Scott 2013). 

This means that observed correlations between dynamically measured BH masses and the galaxy properties automatically select objects for which
\begin{equation}
 \theta_{infl}\equiv r_{infl}/d_{ang} \geq \theta_{crit}
\end{equation}
where $d_{ang}$ is the angular distance of the object and $\theta_{crit}$ is the resolution limit of the instrument (e.g.,  $\theta_{crit}\approx 0.1$ for observations with the {\it Hubble Space Telescope, HST}). 

Shankar et al (2016) have estimated the impact of  such an observational bias on the comparison between the predicted and the observed scaling relations under {\it different assumptions} on the {\it intrinsic} scaling relations. 

Our improved semi-analytic model allows us to  investigate the impact of such an observational bias on the basis of the {\it ab initio} model for galaxy formation presented in the previous section, tested against a wide number of observations.  In fact, our computation of the velocity dispersion $\sigma$  in terms of the properties of the host galaxy enables the prediction of the BH sphere of influence 
$r_{inf}$ for each simulated galaxy of our model. 

In the following we will exploit the model predictions of  $r_{inf}$ to investigate the impact of the selection bias affecting the dynamical BH mass measurements on the comparison between the predicted and the observed BH scaling relations.

\section{Results} 

Before performing a detailed comparison of our predictions for the BH scaling relations with available data, we first show the effects of the model improvements presented in the previous section, i.e., the inclusion of the 2D, angle dependent AGN feedback model presented in Sect. 2.1, and the $\sigma$-dependent selection bias described in Sect. 3. To this aim, 
we compare in fig. 7 the $M_{BH}-M_{*,b}$ relation that we obtain adopting our 2D feedback model (upper panels) with that obtained using the isotropic AGN feedback model previously adopted in our SAM (lower panels). 
To show the effect of the selection bias,  results from models including the bias (right columns) are compared to the ones without such an effect (left columns). 

The comparison between the first and the second column in Fig. 7 shows that that a true relation between the BH mass $M_{BH}$ and the bulge mass is indeed expected in both the isotropic and in the 2D, angle-dependend AGN feedback models, irrespective  of the inclusion of the selection effect resolution of the BH sphere of influence is considered. The main effect of including the selection bias is to extend the distribution of model galaxies toward larger BH masses (especially for large $M_{*,b}$), in better agreement with observations.

It is interesting to note that the slope of the scaling relations in the case of 2D, angle-dependent AGN feedback are almost unchanged in the case of a isotropic AGN feedback. These results lead us to conclude that the detailed form of AGN feedback does not appear as the main ``culprit'' behind the origin of the scaling relations, which seem instead to be mostly driven in our model by the dependence of the BH mass with the processes connected to the growth of the stellar content and of the host galaxy, which in turn imply a correlation between $M_{BH}$ and $\sigma$. Interestingly, Marsden et al. (2022) have recently shown that a steep and nearly constant $M_{BH}$-$\sigma$ relation naturally arises from the coupling of the observed $L_X$-$M_*$ relation and a Jeans-based stellar velocity dispersion (similarly to what carried out here), without any extra fine-tuning.

However, the adoption of the 2D, angle-dependent description of AGN feedback has an important effect on the scatter of the relations, as shown 
quantitatively in the rightmost panels of Fig. 7. The adoption of our new 2D model for AGN feedback reduces 
the intrinsic scatter from values $\epsilon\lesssim 0.5-0.6$ (obtained in  the isotropic feedback case)  to values $\epsilon\lesssim 0.3$. This is due to the following reasons: in the new 2D model for feedback, the blast wave expansions stalls along the direction of the disc, and the radius where the expansion stops depends strongly on both the gas density of the disk and the AGN luminosity. 
This means that the opening angle (and hence the fraction of expelled gas) is larger when the gas density is small (because of the lower energy that has to be spent to push the gas outwards) and when the AGN luminosity is large (because of the larger energy available to push the blast wave outwards), as shown in Fig. 2.
Both quantities depend on the merging histories and are  related, since the AGN luminosity $L_{\rm AGN}$ depends on the available cold gas reservoir $M_{\rm gas}$. The large efficiency of feedback in galaxies with particularly small $M_{\rm gas}$ (for given $L_{\rm AGN}$) or in those with particularly large $L_{\rm AGN}$ (for given $M_{\rm gas}$) inhibits the BH growth in all the host galaxies that are outliers with respect to the average relation between $M_{\rm gas}$ and $L_{\rm AGN}$. This results into a smaller scatter.

\vspace{-0.1cm}
\hspace{-0.cm}
\begin{center}
\scalebox{0.64}[0.64]{
\includegraphics{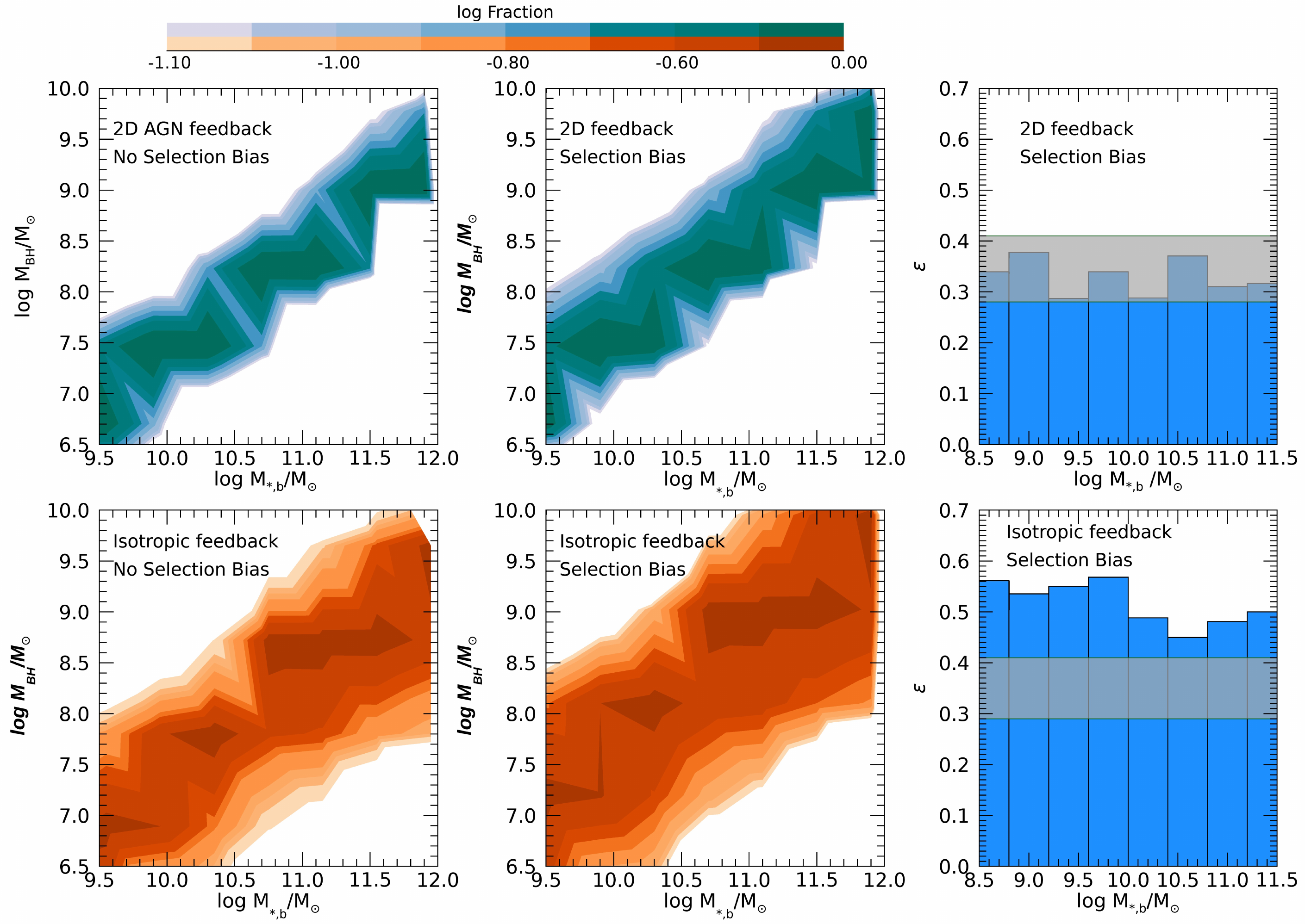}}
\end{center}
\vspace{-0.2cm }
{\footnotesize Fig. 7. 
 The  relation between the BH mass $M_{BH}$ and the bulge stellar mass  $M_{*,b}$, predicted for galaxies at $z\leq 0.1$ by our new 2-D model for AGN feedback (Top Panels) presented in Sect. 2.1 is compared with the predictions obtained with our previous, isotropic treatment of feedback (Bottom Panels). For both AGN feedback models we show the results with and without the inclusion of the selection effect discussed in Sect.2.1, as indicated by the labels. 
In all panels, the color code corresponds to the logarithm of the fraction of galaxies with different $M_{BH}$ in a given log $M_{*,b}$ bin. 
 The standard deviations $\epsilon$ for the new  and the previous treatment of AGN feedback is shown as a function of bulge stellar mass in the rightmost histograms. The horizontal strip represents the range of values obtained from different observational works: the upper bound is the intrinsic scatter $\epsilon$ derived by De Nicola et al. 2021; while the lower bound is the value of $\epsilon$ derived by the least-square fits analysis by  Kormendy \& Ho 2014 using individual errors in log $\sigma$,  adding individual errors in log $M_{BH}$  to the intrinsic scatter in quadrature,  and iterating the intrinsic scatter until the reduced $\chi^{2} = 1$. 
}
\vspace{0.2cm }

The  $M_{BH}-M_{*,b}$ relation predicted by our improved model is compared with data in fig. 8. 
The observational data we compare with include BH estimates for both quiescent and active BHs. 

First, we consider the BH mass measurements
 collected from spatially resolved estimates available from the literature, concerning bulges or bulge-dominated galaxies;   
the data set including bulges and elliptical galaxies in Kormendy \& Ho (2013), and the extended data sets in Savorgan \& Graham (2016) and 
 De Nicola et al. (2019). The first consists of 66 galaxies with dynamical estimates of their black hole masses as reported by Graham \& Scott (2013) or Rusli et al. (2013). Using 3.6 $\mu$m (Spitzer satellite) images, Savorgnan \& Graham (2016) modelled the one-dimensional surface brightness profile of  each of  the 66 considered galaxies and estimated the structural parameters of their spheroidal component by simultaneously fitting a Sérsic function. 
Following Shankar et al. (2016), we retain from the original Savorgnan \& Graham (2016) sample only the galaxies with secure black hole mass measurements and remove those sources classified as ongoing mergers, limiting the final sample to 48 galaxies of which 37 are early-type galaxies (ellipticals or lenticulars), which we consider in our comparison. 
De Nicola et al. (2019), basing on the compilation of Saglia et al. (2016),  derive a sample of 83 galaxies for which BH masses have been derived from spatially resolved kinematics.

We also compare our predicted $M_{BH}-M_{*,b}$ relation with the observed  relation for active galaxies, where the BH masses are derived from the (presumed) virial motions of the Broad Line Region (BLR) gas cloud orbiting in the vicinity of the central compact object:
$ M_{BH}=f_{vir}\,r (\Delta V )^{2}/G$. 
Here $r$ is the radius of the BLR, which is derived from reverberation mapping (e.g., Blandford \& McKee 1982; Peterson 1993), or reverberation-based methods that use the radius–luminosity relation (e.g., Bentz et al. 2006). The characteristic velocity $\Delta V$  is derived from the width of the emission lines (a common one is H$\beta$). As motions in the BLR are not perfectly Keplerian, a parameter $f_{vir}$  is included in the equation to account for the uncertainties in kinematics, geometry, and inclination of the clouds. This is calibrated by comparing reverberation-mapped AGNs with measured bulge stellar velocity dispersions against the $M_{BH}-\sigma$  relation of inactive galaxies. Here we consider the compilation by Ho \& Kim (2014), who derive  a virial coefficient $f_{vir} = 6.3\pm 1.5$ for classical bulges and ellipticals.

 We find that 
 the treatment of the AGN feedback adopted in our semi-analytic model yields BH scaling relations matching the observations when the observational bias related to the resolution of the BH sphere of influence is considered. 
Again,  the inclusion of such a bias extends the distribution of model galaxies toward larger BH masses (especially for large $M_{*,b}$), in better agreement with observations. Shankar et al. 2016 discussed that the degree of the impact of the bias between the normalization of the intrinsic and observed BH-galaxy scaling relations depends on the type of the intrinsic relation one considers. Assuming an intrinsic relation scaling 
$M_{BH}\propto \sigma^{\alpha}$, with $\alpha\gtrsim 4$ would lead to a  ``jump'' between intrinsic and observed relations larger than assuming, for example, $M_{BH}\propto M_*^{\beta}$, with $\beta\approx 1$, as an intrinsic relation. Our current SAM predicts an average difference of $\lesssim 2$ between intrinsic and observed scaling relations, which is what would be expected  if the BH mass has a weaker dependence on velocity dispersion and/or a stronger correlation with stellar mass than with velocity dispersion (Models 2 and 3 in Shankar et al. 2016), as also highlighted in the residuals below.   

\vspace{-0.4cm}
\hspace{0.2cm}
\begin{center}
\scalebox{0.67}[0.67]{\rotatebox{-0}{\includegraphics{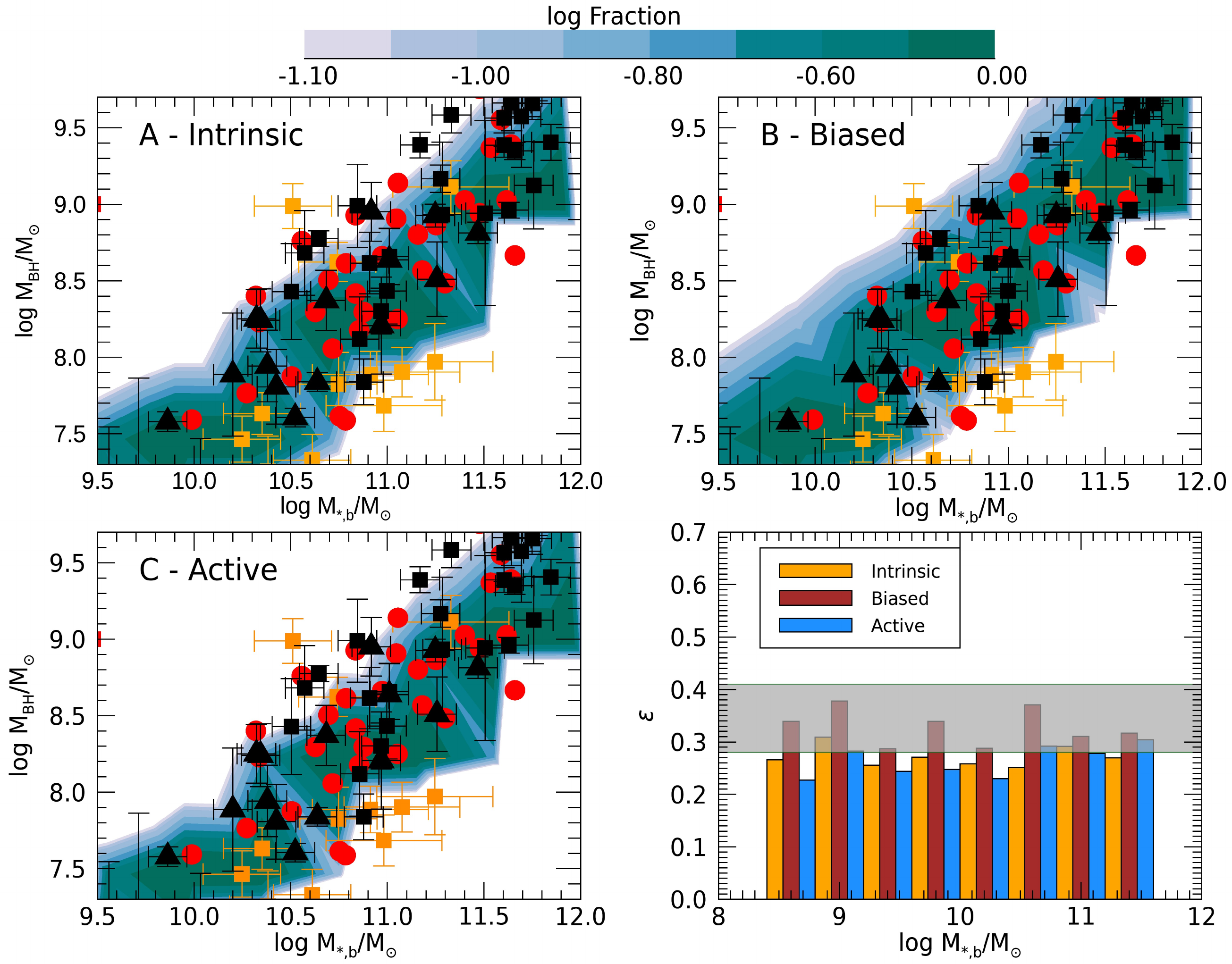}}}
\end{center}
\vspace{0.1cm }
{\footnotesize Fig. 8.  Panel A. The predicted relation between the BH mass $M_{BH}$ and the bulge stellar mass  $M_{*,b}$, for all galaxies at $z\leq 0.1$.
 The color code corresponds to the logarithm of the fraction of galaxies with different $M_{BH}$ in a given log $M_{*,b}$ bin.
The data points are from Kormendy \& Ho 2013 (black squares and triangles for elliptical galaxies and bulges, respectively), Savorgnan \& Graham 2016 (red   dots), and Ho \& Kim (2014, orange filled squares). \newline
Panel B. The same as Panel A but including in the model the selection bias  discussed in section 3, i.e., considering only model galaxies 
satisfying the condition in eq. 12 with $\theta_{crit}=0.1$.\newline
Panel C. The same as panel A but considering only model galaxies hosting AGN with bolometric luminosity $L_{AGN}\ge 10^{44}$ erg/s.\newline
Bottom Right Panel. For different bins of $M_{*,b}$ we show the rms value of the scatter of the distributions shown in Panel A (brown histogram), Panel B (blue histogram), and Panel C (orange  histogram). The horizontal strip corresponds to the range of measured intrinsic scatter by different authors, with the lower value $\epsilon=0.28$ obtained by Kormendy \& Ho (2014) and the upper value $\epsilon=0.4$ obtained by Savorgnan et al (2016). 
}
\vspace{0.2cm }

We also notice that the predicted distribution of model galaxies with an active AGN ($L_{AGN}\ge 10^{44}$ erg/s) does not seems to differ substantially from the distribution of the whole population. However, some of the observational points 
related to active galaxies lie out of the strip representing the model predictions. While a possible explanation is that some of the outflows in these galaxies are spherical, it is possible that the  apparent discrepancy in the normalizations of the $M_{BH}-M_{*,b}$ scaling relations characterizing local AGN data and local inactive BH samples with dynamical mass measurement is due 
to the presence of some additional biases in the BH and/or the stellar mass estimates between the two BH samples, so that the the AGN data represent a 
biased sample of the whole population (see, e.g., Shankar et al. 2019; Farrah et al. 2022, and references therein). We shall further investigate this point below (Fig. 12). 

The scatter we obtain is in good agreement with the observations, for which most observations yield values in the range 
$0.3\lesssim \epsilon\lesssim 0.4$ (see references in the caption), although some authors find somewhat larger values (see, Sahu, Graham, Davies 2019). Indeed, the novel treatment of AGN feedback we consider reduces the scatter of the above relations with respect to the adoption of an isotropic, average feedback. This is shown in Fig. 9, where we compare the $M_{BH}-M_{*,b}$ relation (inclusive of the selection bias described in Sect. 3) obtained with the 2-D model presented in Sect. 2.1 with our previous, isotropic treatment of AGN feedback described in Menci et al. (2014). 

We present in Fig. 9 the predicted relation between  BH mass $M_{BH}$ and the bulge velocity dispersion $\sigma$ for galaxies at $z\leq 0.1$; in addition to the intrinsic relation, we also show the relations that we predict when the selection bias presented in the previous section is considered (i.e., for model galaxies where the condition in eq. 12 is satisfied), and when only active galaxies (i.e., galaxies hosting an AGN with luminosity $L_{bol}\geq 10^{44}$ erg/s) are considered. Although in the plot $\sigma$  refers to the bulge velocity dispersion, we have verified that for galaxies with $B/T\geq 0.3$  the model predictions do not change substantially when the global $\sigma$ is computed from the total galaxy size and total stellar mass $M_*$ (in this case we have used a stellar-mass dependent Sersic index as given in Terzic \& Graham 2005). 

\vspace{-0.1cm}
\hspace{-0.8cm}
\scalebox{0.62}[0.62]{
\includegraphics{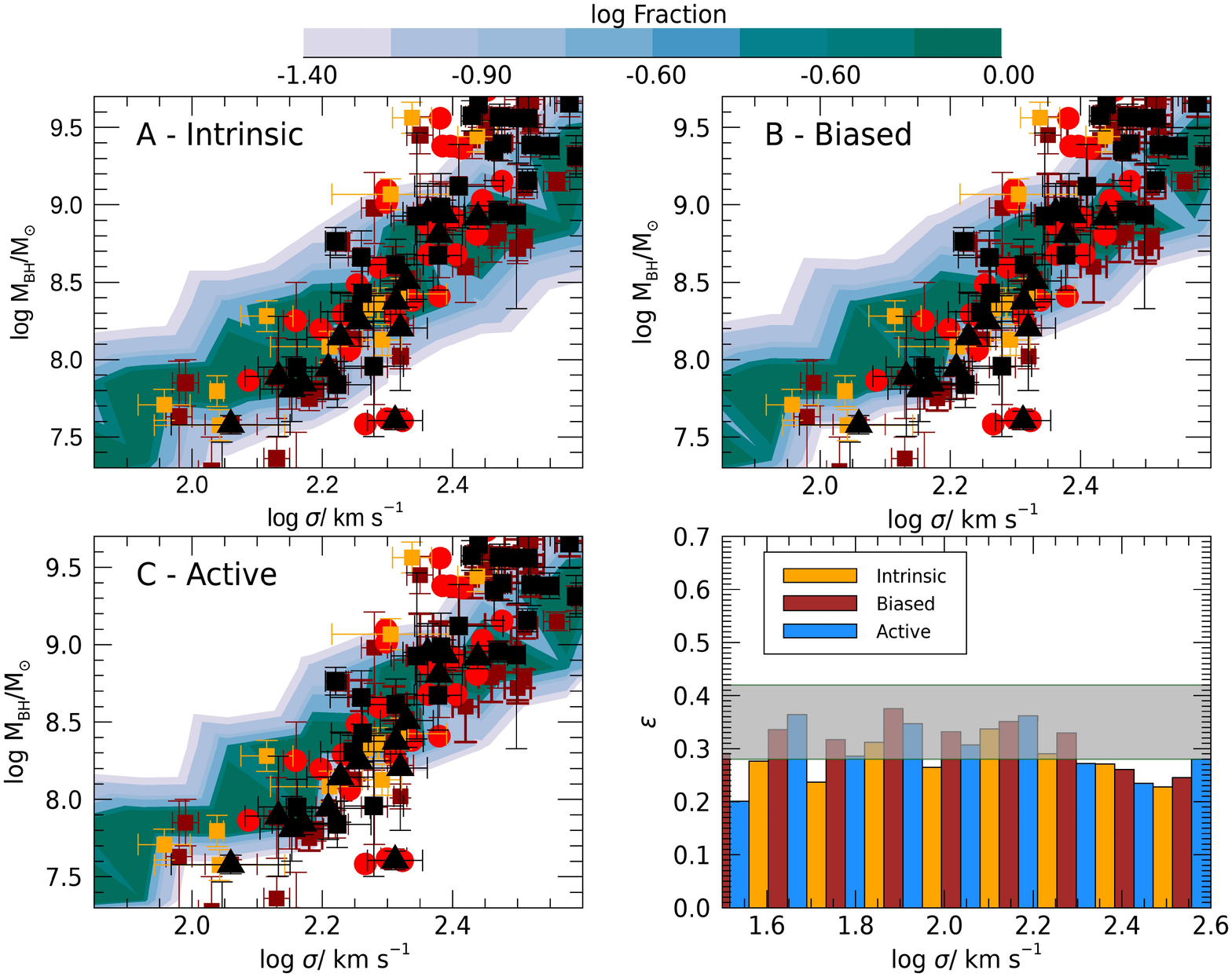}}
\vspace{-0.1cm }
{\footnotesize Fig. 9.  Panel A. The predicted relation between the BH mass $M_{BH}$ and the bulge velocity dispersion $\sigma$  brightness-weighted averaged over the galaxy, for all galaxies at $z\leq 0.1$. 
The color code corresponds to the logarithm of the fraction of galaxies with different $M_{BH}$ in a given log $\sigma$ bin.
The data points are from Kormendy \& Ho 2013 (black squares and triangles for elliptical galaxies and bulges, respectively), De Nicola 2019 (brown squares), Savorgnan e\& Graham 2016 (red dots), and Ho \& Kim (2014, orange filled squares). \newline
Panel B. The same as panel A but including in the model the selection bias  discussed in section 3, i.e., considering only model galaxies 
satisfying the condition in eq. 12 with $\theta_{crit}=0.1$.\newline
Panel C. The same as Panel A but considering only model galaxies hosting AGN with bolometric luminosity $L_{AGN}\ge 10^{44}$ erg/s.\newline 
Bottom Right Panel. The histograms represent the rms values of the scatter for the 
distributions shown in the top left (brown), top right (blue), and bottom left panel (orange). The horizontal strip represents the range of values obtained from different observational works: the upper bound is the intrinsic scatter $\epsilon$ derived by De Nicola et al. 2021; while the lower bound is the value of $\epsilon$ derived by the least-square fits analysis by  Kormendy \& Ho 2014 using individual errors in log $\sigma$,  adding individual errors in log $M_{BH}$  to the intrinsic scatter in quadrature,  and iterating the intrinsic scatter until the reduced $\chi^{2} = 1$. 
}
\vspace{0.2cm }

In the comparison, we assume that stellar mass traces light, since the mass-weighted averaged velocity dispersion in eq. 8 derived for model galaxies is compared with  observed values derived from a brightness-weighted average (see e.g., De Nicola et al. 2019). 

The comparison between the observational data and the model predictions provides interesting insight on some crucial points: \newline
$\bullet$ Comparison between Panels A and B shows that a true relation between the BH mass $M_{BH}$ and the bulge mass is indeed expected, irrespective  
of the inclusion of selection effects.  In the model, the observed correlation does not consist of the upper envelope of the scatter of a distribution of BH masses that  extends to lower BH masses as suggested by some authors (se, e.g., Batcheldor 2010), also in line with the findings by Shankar et al. (2016). \newline
$\bullet$ However, comparing the results in Panel A and B the distribution of  model galaxies in the $M_{BH}-\sigma$ plane is brought to a better 
agreement with the observational data when the selection effect discussed in sect. 3 is considered. In fact, the inclusion of such a bias extends the distribution of model galaxies toward larger BH masses, in better agreement with observations. However, such an extension is less pronounced with respect to that affecting the predicted $M_{BH}-M_{*,b}$ relation. This effect has been already noticed by Shankar et al. (2016), and led these authors to conclude 
that this suggests that the $M_{BH}-
\sigma$ relation is the true fundamental relation, 
while the $M_{BH}-M_{*,b}$  
relation is mostly a consequence of the $M_{BH}-\sigma$ and $\sigma-M_{*,b}$ relation. We will investigate this pointy in detail below.
\newline
$\bullet$ When the selection bias is considered, not only the bulk of the population of model galaxies is consistent the observed relation, but the also the scatter in the distribution is in excellent agreement with the distribution of data points. 
\newline
$\bullet$ It is extremely interesting to note that the SAM predicts a $M_{BH}-\sigma$ relation of similar normalization and dispersion with respect to the collection of available data sets in the local Universe, but with a flatter slope. The predicted $M_{BH}-\sigma$ relation has a slope of $\sim 2.5$ when selection effects are included (panel B), and somewhat lower x ($\sim 2.2$)  in the intrinsic relation. This result is in line with what put forward in other theoretical and numerical works. Cavaliere \& Vittorini (2002; see also Vittorini et al. 2005) highlighted the fact that if the quasar-feedback condition (e.g., Silk \& Rees 1998; King 2003) is not constantly applied throughout the evolution of BHs, the resulting $M_{BH}-\sigma$ relation would end up with a slope closer to $\sim 3$ rather than $4-5$. More recently, Li et al. (2020) showed that the TNG100 simulation also produces a $M_{BH}-\sigma$ relation with a slope of $\sim 3$ (their Figure 1, bottom panels), and in general above the data, where the latter effect could also be, at least in part, ascribed to selection effects in the local observational data sets (e.g., Shankar et al. 2016, Li et al. 2020). Sijacki et al. (2015) and Thomas et al. (2019) found, respectively, in the Illustris and SIMBA simulations, a slope and normalization for the $M_{BH}-\sigma$ relation consistent with observations. However, a $M_{BH} \propto \sigma^\alpha$, with $\alpha \gtrsim 4-5$, in some simulations/models could also be in part an artifact of a steeper $M_*-\sigma$ relation in the same models, as discussed in Barausse et al. (2017), as in the case of the Horizon-AGN simulation (Dubois et al. 2016).
\newline
$\bullet$ Analogously to what we predict for the $M_{BH}-M_{*,b}$ relation,  the predicted distribution of model galaxies with an active AGN ($L_{AGN}\ge 10^{44}$ erg/s) does not seems to differ substantially from the distribution of the whole population. 

In summary, the predicted distribution of galaxies in the $M_{BH}-\sigma$ plane covers the regions allowed by available data. However our SAM , in line with some other state-of-the-art models discussed above, falls short in fully matching the steep slope of the $M_{BH}-\sigma$ relation as inferred from current local data. This could be interpreted in two ways: 1) either the local sample is incomplete, and/or 2) current cosmological models for the coevolution of BHs and galaxies are still missing some ingredients, either in the AGN feedback process itself, and/or in other events during the growth of the BH (e.g., Cavaliere and Vittorini 2002). At our end, the only viable way to increase the slope in the $M_{BH}-\sigma$ relation would be to slightly increase the anisotropy parameter with stellar mass from 0.3 to 0.35, but this would also steepen the $\sigma-M_*$ relation, worsening the match with observations. Of course, a trade-off between the inclusion of the dark matter component and stellar anisotropy could maybe improve the fit to both observables simultaneously. The main lesson we learn from previous studies and the one conducted in this work, is that at face value AGN kinetic feedback, by itself, may not be sufficient to entirely reproduce the current observational data on local BHs, something which is further corroborated by the analysis of the residuals which we present below.  

To go beyond the study of pairwise correlations between  $M_{BH}$, $M_{*,b}$ and $\sigma$, and gain a better insight into the joint distribution of  such quantities, we now focus on  our model’s predictions for the residuals of the BH-galaxy scaling relations. For any couple of quantities $X$ and $Y$, these are defined as $\Delta(Y|X)\equiv log Y-\langle log Y|log X\rangle$. 
As proposed by various authors (see, e.g., Shankar et al. 2016) studying correlations between the residuals provide additional information about the joint distribution of $M_{BH}$, $\sigma$ and $M_*$, and can provide  a way of determining which variable is more important in determining the 
$M_{BH}-M_{*}$ and $M_{BH}-\sigma$ relations.  E.g., in the ideal case where $M_{BH}$ was fundamentally determined by $\sigma$ alone, then residuals from correlations with $\sigma$ would be uncorrelated, while residuals from the $M_{BH}-M_*$ relation would correlate with residuals from the $\sigma-M_*$ relation. 

In the left- and the right-hand panels of Fig. 10 we show 
 $\Delta(M_{BH}|M_{*})$ vs $\Delta(\sigma|M_{*})$ and   $\Delta(M_{BH}|\sigma)$ vs $\Delta(M_{*}|\sigma)$ predicted for 
 bulge-dominated model galaxies  (B/T ratio $\geq 0.8$), and compare such relations with those derived for the Savorgan (2016) sample of early-type galaxies, for which residuals have been computed in Shankar et al. (2016) and Barausse et al. 2017).

The predicted residuals occupy the same regions of the observational data points,  
 showing that the model captures the complex intertwining between the 
 considered quantities. As expected, the predicted behaviour of the residuals is complex, and none of the considered quantities can be considered as really  
 ''fundamental'' in our SAM. In fact, in the framework of cosmological galaxy evolution models like our SAM, the BH mass stems from the complex interplay between the  growth history of halo mass and the physics of baryons, which cannot be easily  reduced to a single fundamental dependency on a single quantity. 

Testing the consistency of such a picture with existing data is not straightforward, due to the fact that the model does not provide definite correlations between the residuals, but rather distributions in the $\Delta(M_{BH}|M_*)-\Delta(\sigma|M_*)$ 
and $\Delta(M_{BH}|\sigma)-\Delta(M_*|\sigma|)$ planes. Thus, we performed a two-dimensional Komlogorov-Smirnov (KS) test to assess the consistency of the predicted distributions with the observational points. To this aim, we follow the approach in Fasano \&  Franceschini (1987) using the extended 2-dimensional KS  distance $D$ defined in Peacock (1983). We run a Monte Carlo simulation to generate synthetic data sets from the predicted distributions, each one with the same number of points $N_{data}$ as the real data set. For each Monte Carlo realization we also  generate $N_{data}$ observational data points distributed according to the measured data points and errorbars.  
 We then computed $D$ for each synthetic data set using the approach mentioned above and count what fraction of the time these synthetic $D$ exceed the $D$ from the real data. The resulting fraction provides the   significance level $Q_{KS}$ for the consistency between the data distribution and the model distribution. For the $\Delta(M_{BH}|\sigma)-\Delta(M_*|\sigma|)$ relation, we obtain a significance level $Q_{KS}=0.24$,  indicating that the former data distribution is compatible with the model predictions. For 
$\Delta(M_{BH}|M_*)-\Delta(\sigma|M_*)$ we obtain  $Q_{KS}=0.1$. 
This value leaves open the possibility that either the sample of local BHs is still  incomplete, and/or that there is still some element missing in the cosmological models that generates the dependence between $M_{BH}$ and $\sigma$ (see also 
the results by Barausse et al. (2017) who claimed an absence of correlation $\Delta(M_{BH}|M_*)$ and $\Delta(\sigma|M_*)$ in both hydrodynamic simulations and SAMs.   
 
\vspace{-0.4cm}
\hspace{0.2cm}
\begin{center}
\scalebox{0.64}[0.64]{\rotatebox{-0}{\includegraphics{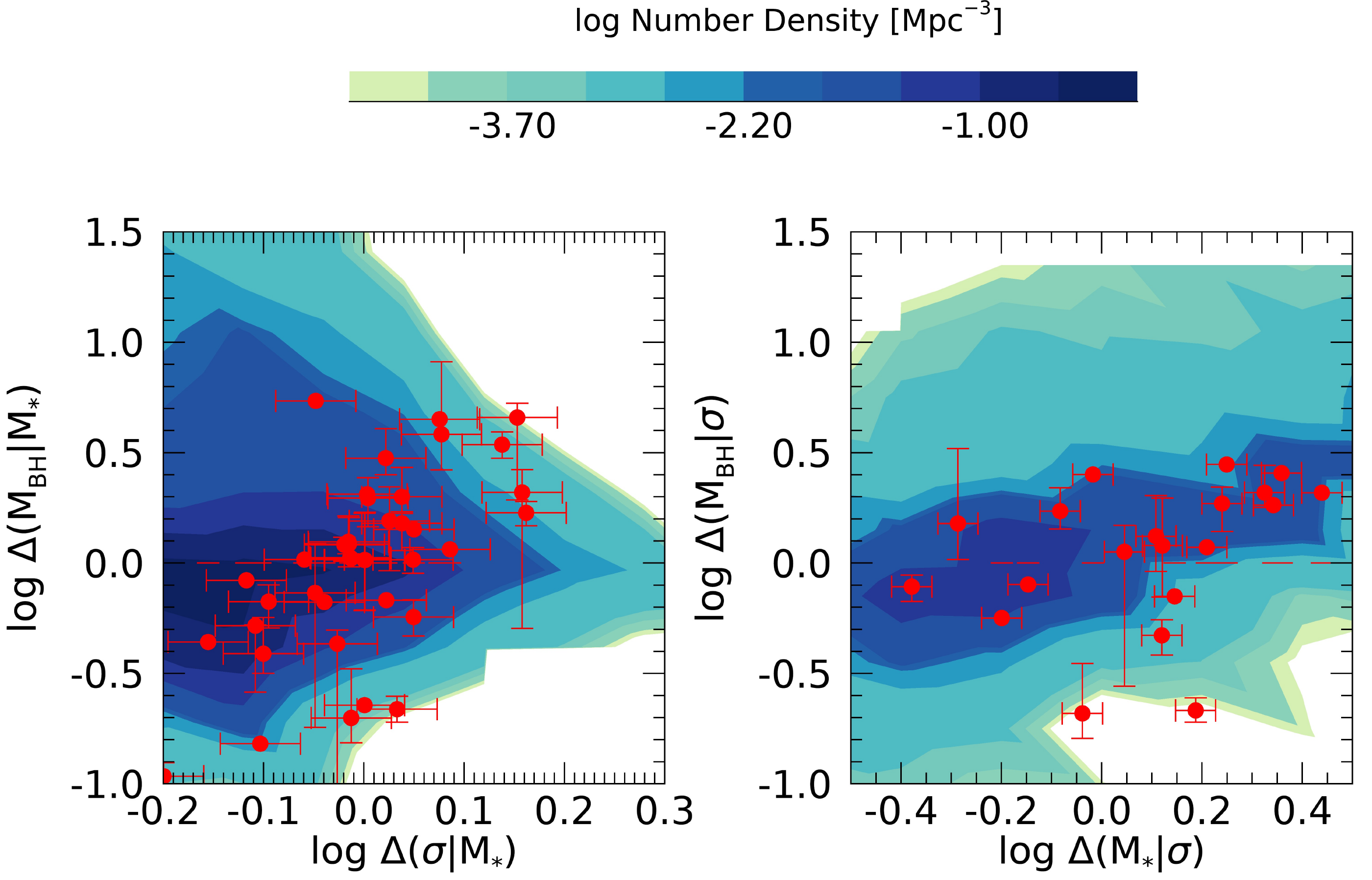}}}
\end{center}
\vspace{-0.3cm }
{\footnotesize Fig. 10. 
Correlation between the residuals of the $M_{BH}-M_{*,b}$ relations and those of the $\sigma–M_*$ relation, at fixed stellar mass (left-hand panel), and between the residuals of the $M_{BH}-\sigma$ relation and those of the $M_*-\sigma$  relation, at fixed velocity dispersion (right-hand panel). Data points refer to Savorgnan et al. (2016) sample. The  contours refer to model galaxies with 
bulge-to-total stellar mass ratio $B/T > 0.8$ for which the black hole sphere of influence is resolvable.
}
\vspace{0.2cm}

Indeed, in such a context we expect BH masses to be correlated with different DM and baryonic properties of galaxies, all emerging from the above complex interplay. To investigate this point in closer detail, we compute the relation between the BH mass and the global properties of galaxies, like the total stellar mass content and the DM mass. We stress that  such relations are expected 
in {\it all} cosmological models of galaxy formation, since 
all baryonic processes (gas cooling and accretion, inflows, star formation and feedback)  are ultimately related to the depth and growth history of the gravitational potential wells. In this framework, the growth of the stellar, BH, and DM is intertwined so that $M_{BH}$, $M_*$, and $M_h$ are expected to be related to each other. The connection among these quantities is investigated in Fig. 11, and compared with recent observational measurements by Marasco et al, (2021). These authors identified a sample of 55 nearby galaxies with dynamically measured $M_{BH} \geq  10^{6}\,M_{\odot}$ for which dynamical measurements of $M_h$ were available, based either on globular cluster dynamics (for galaxies of earlier Hubble types) or on spatially resolved rotation curves (for galaxies of later Hubble types). 
The correlation we find between the BH mass $M_{BH}$ and the total stellar mass content $M_*$ is in excellent agreement with observations in both strengths and scatter, the later being characterized by an intrinsic value $\epsilon \approx 0.45$.  As expected, we find an extremely well defined correlation between $M_{BH}$ and $M_h$, in agreement with the results in Marasco et al. (2021) and with previous authors (Ferrarese  2002; Pizzella et al. 2005; Volonteri, Natarajan \& Gultekin 2011; Davis, Graham, Combes 2019; Shankar et al. 2020; Smith et al.  2021) who used   the  $v_{rot}$  as a proxy for the DM mass $M_{h}$. 
Although our model includes as a key process the merging of BHs, our predicted $M_{BH}-M_*$ relation does not show any sign of flattening at high masses, a fature that has been suggested as indicative of merging events as the leading process in the building up of the relation for early type, massive galaxies
(see Graham, Sahu 2023). This suggest that the gas physics is relavant in the building up of the relation even for massive galaxies.

The tightness of the relation is directly related to the presence of AGN feedback. Indeed, in a model where no AGN feedback is implemented, the relation between $M_{BH}$ and  $M_*$ is expected to be much broader, as shown by the grey contours in Panel A of Fig. 11. 

Thus, the observed tight correlation between $M_{BH}$ and $M_*$ results from the interplay between the growth of DM halos (and the ensuing collapse and the cooling of gas in the DM potential wells, and the related dynamical processes affecting the distribution of gas), and the counteracting effects of AGN outflows which keeps the relation tight. 

\begin{center}
\scalebox{0.82}[0.82]{\rotatebox{0}{\includegraphics{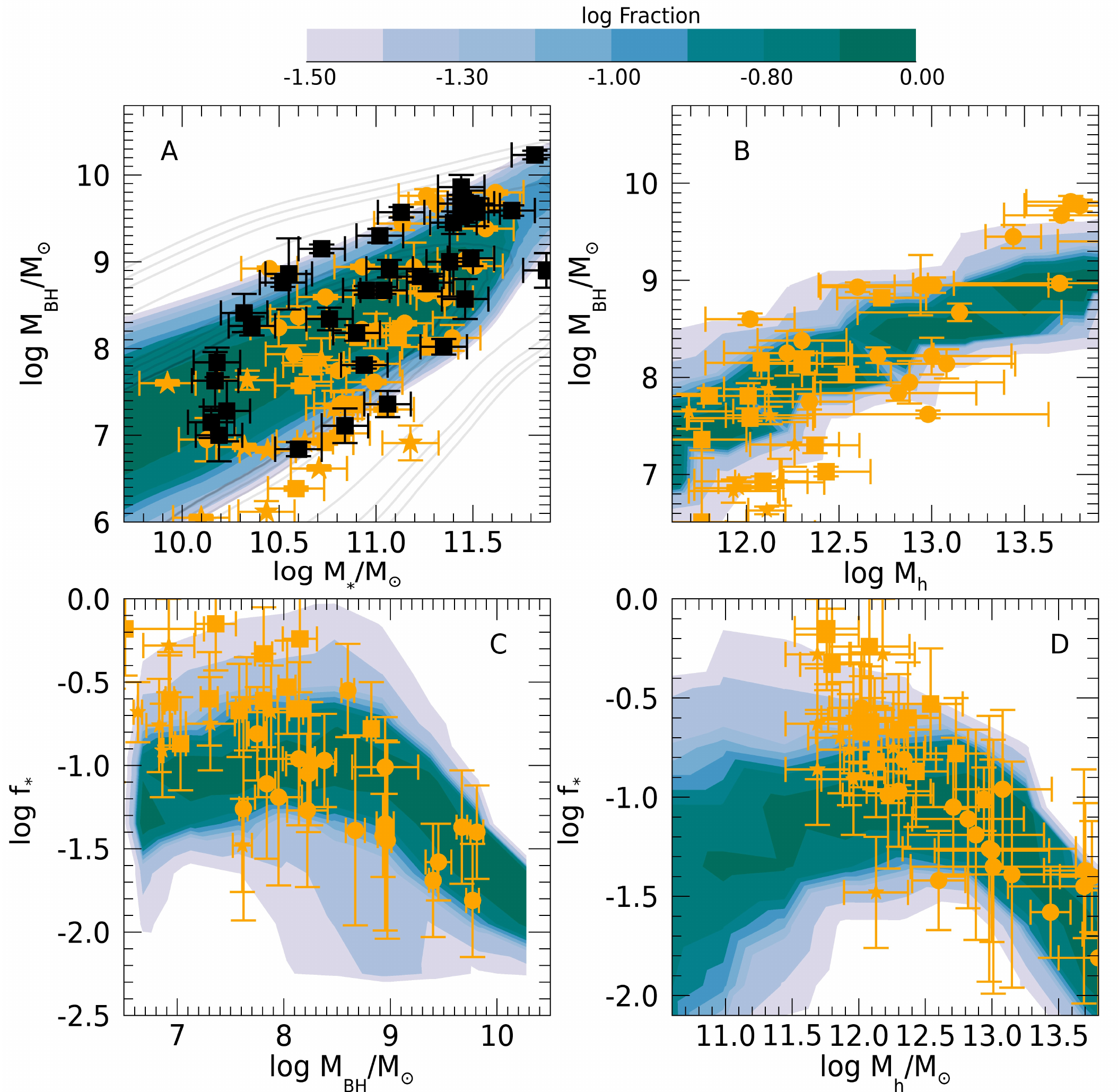}}}
\end{center}
\vspace{0.cm }
{\footnotesize Fig. 11. The predicted $M_{BH}$-$M*$ relation (Panel A), the  $M_{BH}$-$M_{h}$ relation (Panel B), the $f_*$-$M_{BH}$ relation (Panel C) and the $f_*$-$M_{h}$ relation (Panel D) are compared with the data from 
Marasco et al. 2021 (orange points). In Panel A we also report data from Sahu, Graham, Davies 2019 (black dots); the grey contours show the predictions of our model in the absence of AGN feedback. 
In all panels, for each abscissa value, the color code corresponds to the logarithm of the fraction of galaxies with different values of the quantities in the y-axis. }

\vspace{0.2cm }

Since all baryonic processes entering the BH growth are ultimately related to the growth of the DM halos, we expect a tight correlation between the BH mass $M_{BH}$ and the DM halo mass of the host $M_{H}$. This is shown in Panel B, and compared with data. Indeed the relations resulting from the model looks tighter than the data distribution. 
However, the large errorbars of the data points at present do not allow to validate or falsify the model prediction. Here we note that recent observations of central galaxies in galaxy clusters show an extremely tight 
relation between the BH  and the halo mass (Gasparri et al. 2019), indeed tighter than the $M_{BH}$-$M*$ relations, supporting our prediction of a strong correlation between $M_{BH}$ and $M_h$, as found in
 {\it ab initio} models and cosmological simulations (see Davies, Pontzen, Crain 2023 for recent results,  and references therein). 

 However, as stressed by  the authors above, complex physics of baryons in the growing DM potential wells results into a number of correlations. 
Like all cosmological models of galaxy formation, our SAM predicts a strict relation between  the depth of the gravitational potential wells and the star formation processes (Panel D). In particular, the star formation efficiency $f_*\equiv M_*/f_b\,M_h$ (here $f_b$ is the Universal baryon fraction) 
is predicted to have a maximum for DM masses around $M_h\approx 10^{12}\,M_{\odot}$. In fact, for smaller values of $M_h$ star formation is suppressed by 
 the strong effects of stellar feedback, which effectively expels a relevant fraction of gas from the shallow potential wells, while in massive haloes with $M_h\gtrsim 10^{12}\,M_{\odot}$ inefficiency of gas cooling and strong AGN winds combine to quench star formation leading to a declining $f_*$ (see also Shankar et al. 2006). Thus, matching the observed behaviour (Panel D) provides a key test for our implemented 2-dimensional AGN feedback model. It is important to stress that such a model provides a {\it combined} correlation between the quenching properties of AGN feedback 
 (the quantity $f_*$) and its function of self-regulation in the growth of BHs, as shown in Panel C. The consistency between the predicted and the observed relation between $M_{BH}$ and $f_*$ provides an important consistency check 
 for our implemented AGN feedback model, since it provides a simultaneous quantitative description of both the star formation quenching  and the regulation of the BH growth that is consistent with available measurements. 

\begin{center}
\scalebox{0.65}[0.65]{\rotatebox{0}{\includegraphics{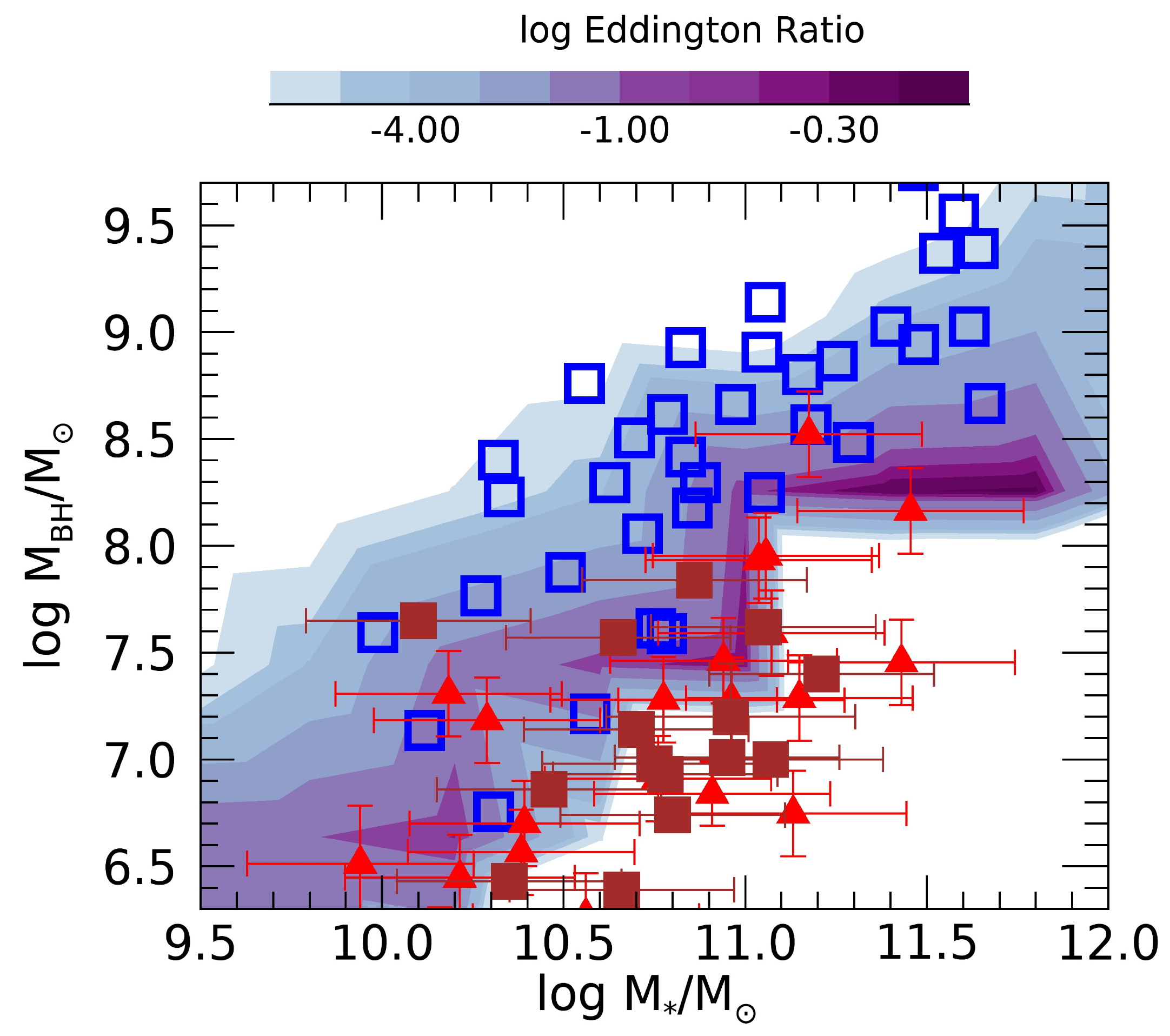}}}
\end{center}
\vspace{0.cm }
{\footnotesize Fig. 12. The color code shows  
the average Eddington ratio of BHs in the  $M_{BH}$-$M_*$ plane. The blue data are taken from Savorgnan \& Graham (2016), while the red squares and triangles are the data obtained from reverberation mapping and maser techniques, respectively. These are taken from van den Bosch et al. (2016). To the errorbars reported by the latter authors (referring to measurement errors in the K-band magnitude) we have added (in quadrature) a systematic error estimate of 0.3 dex  corresponding to the uncertainties in the conversion from k-band magnitudes to stellar masses.
}
\vspace{0.2cm }

Such a correlation between the AGN feedback and the star formation properties of galaxies is expected to impact over the distribution of BHs with different degree of activity in the $M_{BH}$-$M_*$ plane. This is investigated in detail in Fig. 12, where 
we show the  $M_{BH}$-$M*$ relation for different BH Eddington ratios $\lambda$, and compare with data concerning active AGNs with BH masses measured through either  
reverberation mapping and maser techniques (van den Bosch et al. 2016), so that they are not affected by uncertainties related to $f_{vir}$. As noted by previous authors (e.g., Reines \& Volonteri 2015; Shankar et al. 2019) these objects appear to lie below the relation for 
the global BH population. Interestingly, although the model predicts BH masses of active galaxies somewhat larger than the observed, it predicts objects with larger 
Eddington ratios to lie systematically along the lower envelope of the global $M_{BH}$-$M*$ relation. The physical interpretation of this trend is that extremely large accretion rates only take place in objects where large gas reservoirs are still available, and have not been converted into stars and large BH masses at higher redshifts. Although the distribution is consistent with available data, we notice that a detailed verification of our prediction would require a larger collection of  observational points which would enable to sample the distribution of Eddington ratios with sufficient statistically significance. The apparent discrepancy/offset in the normalizations of the $M_{BH}$-$M*$ scaling relations characterizing local AGN data and local inactive BH samples with dynamical mass measurements, is a matter of intense debate (e.g., Reines \& Volonteri 2015; Shankar et al. 2019; and references therein), and not yet solved. Although it is beyond the scope of the current work to deepen into this tension, we note that the dispersion predicted by our SAM is more contained than the one observed in the combined data sets, possibly suggesting the presence of some additional biases in the BH and/or the stellar mass estimates between the two BH samples.

Finally, we investigate the predictions of our updated model on the evolution of the $M_{BH}-M_*$ relation. Specifically, we compare with the recent measurements at $1.2\leq z\leq 1.7$ based on 32 X-ray- selected, broad-line AGN performed by Ding et al. (2020).  By applying state-of-the-art tools to decompose the HST images including available ACS data, the authors  measured the host galaxy  stellar mass  through the two-dimensional model fitting, while the BH mass, was determined using the broad H$\alpha$ line, detected in the near-infrared with the Subaru Fiber Multi-Object Spectrograph. 

To ensure a fair comparison with the above observations, we reproduce the selection biases when considering the model galaxies. Specifically, we adopt the same selection window adopted in the observations $\lambda_{Edd}\geq -0.5-0.11*(log (M_{BH}/M_{\odot})-7.5$, based  on the Eddington ratio $\lambda_{Edd}$ (see Fig. 1 of Ding et al. 2020). 
In addition, following Ding et al. (2020), we injected random noise to the sample of model galaxies to mimic the scatter in the data due to measurement errors, i.e., $\delta M_{BH}= 0.4$ dex,  and $\delta M_* = 0.17$ dex, so as to allow for a quantitative comparison between the observed and predicted scatter in the $M_{BH}-M_*$ relations. 

The results of the comparison is shown in Fig. 13, where we show the prediction for both the whole sample of model galaxies and for the sub-sample selected so as to mimic the observational measurement errors and selection effects. When the  
latter are considered, the distribution of model galaxies matches the observations in both central value and scatter. The former, 
shows a small positive evolution $\Delta M_{BH}\approx 0.5$ with  respect to the local relation, which is consistent with results from N-body simulation (see, e.g., Sijacki et al. 2014; Khandai et al. 2014; De Graf et al. 2015). On the other hand, the scatter 
 takes values ranging from $\epsilon\approx 0.3$ to  $\epsilon\approx 0.38$ , in excellent agreement with the observed value $\epsilon \approx 0.35$. 

\vspace{-0.2cm}
\hspace{-0.8cm}
\begin{center}
\scalebox{0.58}[0.58]{\rotatebox{0}{\includegraphics{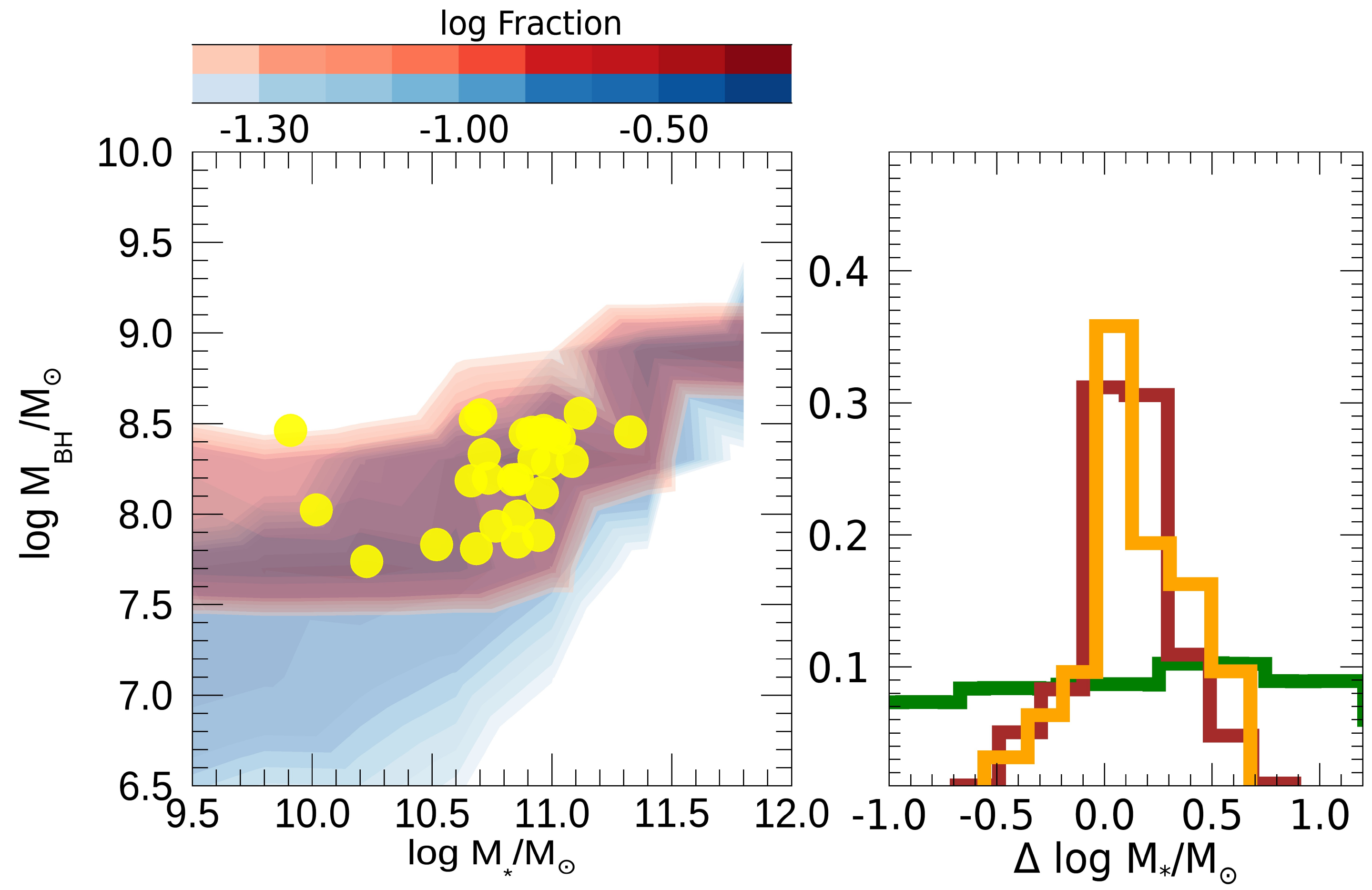}}}
\end{center}
\vspace{0.2cm }
{\footnotesize Fig. 13.  Left Panel. Comparison of the observed (orange dots) and predicted  $M_{BH}–M_*$ relation for galaxies in the redshift range $1.2\leq z\leq 1.7$. 
The color code corresponds to the logarithm of the fraction of galaxies with different $M_{BH}$ in a given log $M_*$ bin.
The data are taken from Ding et al. 2020. The total predicted distribution of 
model galaxies in the $M_{BH}-M_*$ plane is shown by the blue contours, while the red region corresponds to the sub-sample of model galaxies which satisfies the  same selection criteria of the observations. \newline
Right Panel. The distribution of the residuals for the observations (orange histogram), the model galaxies (blue histogram), and the model galaxies resulting in simulations 
with the isotropic feedback model implemented in previous versions of our SAM (green histogram, taken from Ding et al. 2020). Model galaxies are selected  so as to mimic the observational measurement errors and selection effects (see text). 
}
\vspace{0.2cm }

The latter constitutes a remarkable results, which can be entirely ascribed to our improved description of AGN feedback. In fact, the same comparison with observations was performed in Ding et al. (2020) based on the SAM with the  simplified, isotropic description of AGN feedback, finding a much larger 
 scatter $\epsilon \approx 0.7$.  The distribution of the residual based on $M_*$ (i.e., along the x-axis) $\Delta (M_{*}$  is shown on the right of Fig. 12, where the observed distribution is compared with the predictions from our SAM with the new  2-dimensional treatment of feedback (Sect. 2.1), 
 and with those referring to our SAM with the previous isotropic treatment of AGN feedback, showing how the latter are characterized by a much wider distribution of residuals. Thus, our comparative test based on one numerical model and altering the AGN feedback prescription enlightens the role of feedback 
 in regulating the scatter of the observed relations not only at low redshift (Fig. 9) but also at higher redshifts $z\approx 1.5$. 
 
\section{Summary and Conclusions} 

The true role of AGN feedback in shaping galaxies is still a matter of intense debate both observationally and theoretically. To tackle this fundamental problem, we implemented a new physical treatment of AGN-driven winds into our semi-analytic model of galaxy formation. This is based on a two-dimensional description for the expansion of AGN-driven shocks in realistic galactic gaseous discs, which has been successfully tested against a wide set of observed outflows (Menci et al. 2019).
To each galaxy in our model, we associate solutions for the outflow expansion and the mass outflow rates in different directions with respect to the plane of the disc (Fig. 1), depending on the AGN luminosity and on the circular velocity and gas content of the considered galaxy Figs. 2-3).
In addition, we have also updated our semi-analytic model by self-consistently computing for each galaxy in our SAM the stellar velocity dispersion via detailed Jeans modelling. 

We applied our SAM to the study of the scaling relations between the mass $M_{BH}$ of central Black Holes and several properties of galactic bulges, like the stellar mass and the velocity dispersion of the bulge, and  some global properties of the model galaxies, like the total stellar content,  the dark matter mass, and the star formation efficiency. Our main results can be summarized as follows: 

$\bullet$ The implementation in the SAM of the observational limited-resolution effects 
allows us to confirm that a true correlation between the BH mass $M_{BH}$ and galaxy properties (bulge mass and velocity dispersion, total stellar mass, DM mass) do exist (Figs. 8-9), and that the observed relations do not consist of the upper envelope of the scatter of a distribution of BH masses that  extends to lower BH masses as suggested by some authors (see, e.g., Batcheldor 2010). In this context, the model results are consistent with the description of BH growth that emerges from 
hydrodynamical N-body simulations, like, e.g.,  the 
 Magneticum Pathfinder SPH simulations (Steinborn et al. 2015), the Evolution and Assembly of Galaxies and their Environments suite of SPH simulations (Schaye et al. 2015), the Illustris moving mesh simulation (Vogelsberger et al. 2014; Sijacki et al. 2015; Li et al. 2020) and the SIMBA simulation (Thomas et al. 2019).

$\bullet$ When the above limited-resolution selection bias is included, the model predictions yield a  normalization of the  relations  slightly increased (by a factor $\approx 1.5-2$)  toward larger BH masses as expected, bringing them in better agreement with observations in both normalization and scatter (Figs. 7-8). 

$\bullet$ Our predicted BH-galaxy scaling relations are flatter than the observed ones, in particular the $M_{BH}$-$\sigma$ relation, despite the adoption of an improved kinetic AGN feedback model {Fig. 8). Our current work thus does not support the view for AGN feedback being the main driver behind the origin of a steep $M_{BH}$-$\sigma$ relation.

$\bullet$ Compared to the isotropic treatment of AGN feedback implemented in the previous version of our SAM, the effect of a 2-dimensional treatment of AGN outflows 
 in realistic galactic discs is to reduce the scatter of the observed relations (Fig. 7). Thus a reduction (from $\epsilon\approx 0.55$ to $\epsilon \approx 0.3$ in the local Universe) is larger at high redshifts (from $\epsilon\approx 0.75$ to $\epsilon \approx 0.3$). The resulting scatter of the $M_{BH}$-$M_*$ relation is approximately constant out to $z\approx 1.5$. 
 
$\bullet$ We predict a small positive evolution $\Delta M_{BH}\approx 0.5$ in the normalization of the average $M_{BH}$-$M_*$ relation at $z\approx 1.5$ with  respect to the local relation (Fig. 13), which is consistent with results from N-body simulation (see, e.g., Sijacki et al. 2014; Khandai et al. 2015; De Graf et al. 2015).

$\bullet$ 
In our SAM, none of the considered galactic properties  can be considered as really  ''fundamental'' in determining the observed scaling relations, not even stellar velocity dispersion. In fact, in the framework of cosmological galaxy evolution models like our SAM, the BH mass stems from the complex interplay between the  growth history of halo mass and the physics of baryons, which cannot be reduced to a single fundamental dependency on a single quantity. To test such a prediction, we have analyzed the residuals of the scaling relations (Fig. 10), and performed a Kolmogorov-Smirnov test to probe the consistency of their distribution with the distribution of data points. For the $\Delta(M_{BH}|\sigma)-\Delta(M_*|\sigma|)$ relation, we obtain a significance level $Q_{KS}=0.24$, indicating that the former data distribution is compatible with the model predictions, while for 
$\Delta(M_{BH}|M_*)-\Delta(\sigma|M_*)$ we obtain  $Q_{KS}=0.1$. 
This value leaves open the possibility that either the sample of local BHs is still  incomplete, and/or that there is still some element missing in the cosmological models that generates the dependence between $M_{BH}$ and $\sigma$.

$\bullet$ The model predicts a correlation between the BH mass $M_{BH}$ and the total DM mass $M_h$ of the host galaxy which is consistent with available data (Fig. 11, panel B), with a small scatter that is comparable to the scatter of the correlations with $M_*$ or $\sigma$. As found in several N-body simulations (see, e.g., Davies, Pontzen \& Crain 2023)
 the main trigger for BH growth is constituted by galaxy merging. Indeed, switching off disc instabilities in our model does not affect appreciably the correlations we find between the BH mass and the galaxy properties. However, we do not find breaks in the $M_{BH}$.

$\bullet$ The implemented AGN feedback model provides a quenching of star formation in massive halos (Fig. 11, panel D) that is consistent with present observations. 

$\bullet$  Our treatment of the AGN feedback allows our SAM not only to yield  correlations between the BH mass and both $M_*$ and $M_h$ separately,  but also 
to provide relations with the efficiency of star formation in different DM halos  $f_*\equiv M_*/f_b\,M_h$ (Fig. 11, panel C). Thus, it provides a simultaneous quantitative description of both the star formation quenching  and the regulation of the BH growth that is consistent with available measurements. 

The implementation of the 2-dimensional treatment of AGN feedback in our model opens 
the way to several additional studies. For instance, the physical treatment of outflows in a realistic 2-D geometry allows us to compute the escape fraction of ionizing photons associated to each galaxy of the model. This will be tightly related to the opening angle (Fig. 2) computed for each combination of galaxy quantities ($M_{gas}$, $L_{AGN}$ and $M_h$). We plan to investigate this issue in a future paper, and to address the contribution of AGN to the reionization of the Universe (see Giallongo et al. 2012) based on a more realistic and reliable description of AGN outflows.

\begin{acknowledgements}
We thank the referee for helpful comments. We acknowledge grants from MIUR (PRIN MIUR contract 2017PH3WAT). FS acknowledges partial support from the European Union's Horizon 2020 research and innovation programme under the Marie Skłodowska-Curie grant agreement No. 860744.
\end{acknowledgements}


\begin{thebibliography}{}
\bibitem{}Alexander, D. M., Hickox, R. C., 2012, New Astron. Rev., 56, 93
\bibitem{}Allevato, V., Shankar, F., Marsden, C., et al.\ 2021, ApJ, 916, 34 
\bibitem{}Baes, M., Buyle, P., Hau, G.~K.~T., et al.\ 2003, MNRAS, 341, L44 
\bibitem{}Balogh, M. L., Morris, S. L., Yee, H. K. C., Carlberg, R. G., Ellingson, E., 1999, ApJ, 527, 54
\bibitem{}Bandara, K., Crampton, D., \& Simard, L.\ 2009, ApJ, 704, 1135 
\bibitem{}Barausse, E., Shankar, F., Bernardi, M., 2017, MNRAS, 468, 4782
\bibitem{}Batcheldor, D., 2010, ApJ, 711, L108
\bibitem{}Belli, S., Newman, A. B., Ellis, R. S., 2014, ApJ, 783, 117
\bibitem{}Bentz, M. C., Peterson, B. M., Pogge, R. W., Vestergaard, M., Onken, C. A., 2006, ApJ, 644, 133
\bibitem{}Bentz, M.C., Manne-Nicholas, E., 2018, ApJ, 864, 146
\bibitem{}Bernardi, M., Sheth, R.~K., Tundo, E., et al.\ 2007, ApJ, 660, 267
\bibitem{}Bernardi, M., Roche, N., Shankar, F., Sheth, R.K., 2011, MNRAS, 412, L6
\bibitem{}Bernardi M., Meert A., Vikram V., Huertas-Company M., Mei S., Shankar F., Sheth R. K., 2014, MNRAS, 443, 874
\bibitem{}Blandford R. D., McKee C. F., 1982, ApJ, 255, 419
\bibitem{}Bond, J. R., Cole, S., Efstathiou, G., Kaiser, N. 1991, ApJ, 379, 440
\bibitem{}Boutsia, K. et al. 2018, ApJ, 869, 20
\bibitem{}Boutsia, K. et al. 2021, ApJ, 912, 111
\bibitem{}Burkert, A. \& Tremaine, S.\ 2010, ApJ, 720, 516 
\bibitem{}Busch G. et al., 2014, A\&A, 561, A140
\bibitem{}Cavaliere, A., Lapi, A.,  Menci, N. 2002, ApJ, 581, L1
\bibitem{}Cavaliere A., Vittorini V., 2002, ApJ, 570, 114
\bibitem{}Cicone, C., Maiolino, R., Gallerani, S., et al. 2014, A\&A, 562, 21
\bibitem{}Cirasuolo, M., Shankar, F., Granato, G.~L., et al.\ 2005, ApJ, 629, 816
\bibitem{}Cole S., Lacey C. G., Baugh C. M., Frenk C. S., 2000, MNRAS, 319, 168
\bibitem{}Davis B. L., Graham A. W., Cameron E., 2019, ApJ, 873, 85
\bibitem{}Davis B. L., Graham A. W., Combes F., 2019, ApJ, 877, 64
\bibitem{}Davies, J.J., Pontzen, A., Crain, R.A. 2023, preprint [arXiv:2301.04145]
\bibitem{}De Graf, C., Di Matteo, T., Treu, T., et al. 2015, MNRAS, 454, 913
\bibitem{}Dekel, A., Silk, J., 1986, ApJ, 303, 39
\bibitem{}Desmond H., Wechsler R. H., 2017, MNRAS, 465, 820
\bibitem{}De Nicola, S., Marconi, A., Longo, G. 2019, MNRAS, 490, 600
\bibitem{}Ding, X., Treu, T. Silverman, J.H. et al. 2020, ApJ, 896, 159
\bibitem{}Efstathiou G., Lake G., Negroponte J., 1982, MNRAS, 199, 1069
\bibitem{}Fanidakis, N. et al. 2012, MNRAS, 419, 2797
\bibitem{}Farrah, D., Petty, S., Croker, K., et al.\ 2022, arXiv:2212.06854
\bibitem{}Faucher-Giguere C.-A., Quataert E., 2012, MNRAS, 425, 605
\bibitem{}Fasan, G., Franceschini, A., 1987, MNRAS, 225, 155 
\bibitem{}Ferrarese, L., 2002, ApJ, 578, 90
\bibitem{}Ferrarese, L., Ford, H. 2005, SSRev, 116, 523
\bibitem{}Fiore, F., Feruglio, C., Shankar, F., et al. 2017, A\&A, 601, A143
\bibitem{}Fontanot, F., Monaco, P., \& Shankar, F.\ 2015, MNRAS, 453, 4112
\bibitem{}Fontanot, F., De Lucia, G., Hirschmann, M., et al.\ 2020, MNRAS, 496, 3943
\bibitem{}Gasparri, M. et al. 2019, ApJ, 884, 169
\bibitem{}Gebhardt, K., Bender, R., Bower, G., et al.\ 2000, ApJL, 539, L13
\bibitem{}Genel, S. et al., 2018, MNRAS, 474, 3976
\bibitem{}Giallongo, E., Menci, N., Fiore, F., et al. 2012, ApJ, 755, 124
\bibitem{}Giallongo, E., Grazian, A., Fiore, F., et al. 2019, ApJ, 884, 19
\bibitem{}Graham, A. W., Scott N., 2013, ApJ, 764, 151
\bibitem{}Graham, A. W., 2016, in Laurikainen E., Peletier R., Gadotti D., eds, Galaxy
Bulges. Springer Int. Publ., Cham, p. 263
\bibitem{}Graham, A.W., 2022, MNRAS, in press, arXiv:2211.02187
\bibitem{}Graham, A.W., Sahu, N. 2023, MNRAS,  518, 2177
\bibitem{}Granato, G.L., De Zotti, G., Silva, L. Bressan, A., Danese, L. 2004, ApJ, 600, 580
\bibitem{}Greene J. E. et al., 2016, ApJ, 826, L32
\bibitem{}Grazian, A., Giallongo, E., Fiore, F., et al. 2020, ApJ, 897, 94
\bibitem{}G\"ultekin, K., Tremaine, S., Loeb, A., Richstone, D. O., 2011, ApJ, 738, 17
\bibitem{}Guo, Q., White, S., Boylan-Kolchin, M., et al.\ 2011, MNRAS, 413, 101
\bibitem{}Habouzit, M., Li, Y., Somerville, R.~S., et al.\ 2021, MNRAS, 503, 1940
\bibitem{}Hartwig, T., Volonteri, M., Dashyan, G., 2018, MNRAS, 476, 2288
\bibitem{}Hirschmann, M., Somerville, R.S., Naab, T., Burkert, A., 2012, MNRAS, 426, 237
\bibitem{}Hyde, J.B., Bernardi, M., 2009, MNRAS, 394, 1978
\bibitem{}Ho L. C., Kim M., 2014, ApJ, 789, 17
\bibitem{}Hopkins, P. F., Hernquist, L., Cox, T. J., Matteo, T. D., Robertson, B., Springel, V., 2006, ApJS, 163, 1
\bibitem{}Hopkins, P.~F., Hernquist, L., Cox, T.~J., et al.\ 2007, ApJ, 669, 67
\bibitem{}Hopkins P. F., Hernquist L., Cox T. J., Keres D., Wuyts S., 2009, ApJ, 691, 1424
\bibitem{}Hyde J. B., Bernardi M., 2009, MNRAS, 394, 1978
\bibitem{}Iannella, A.~L., Greco, L., \& Feoli, A., 2021, ApSS, 366, 52
\bibitem{}Khandai, N., Di Matteo, T., Croft, R., et al. 2015, MNRAS, 450, 1349
\bibitem{}King A., 2003, ApJ, 596, L27
\bibitem{}King A. R., Zubovas K., Power C., 2011, MNRAS, 415, L6
\bibitem{}King A., Pounds K., 2015, ARAA, 53, 115
\bibitem{}Kormendy, J., Gebhardt, K. 2001, in 20th Texas Symposium on Relativistic Astrophysics, ed. J. C. Wheeler,  H. Martel (Melville, NY: AIP), 363
\bibitem{}Kormendy J., Bender R., 2011, Nature, 469, 377
\bibitem{}Kormendy J., Ho L. C., 2013, ARA\&A, 51, 511
\bibitem{}Lacey, C., Cole, S., 1993, MNRAS, 262, 627
\bibitem{}Lapi, A., Cavaliere, A., Menci, N. 2005, ApJ, 619, 60
\bibitem{}Lapi, A., Shankar, F., Mao, J., Granato, G. L., Silva, L., Zotti, G. D., Danese L., 2006, ApJ, 650, 42
\bibitem{}Li, Y. et al., 2020, ApJ, 895, 102
\bibitem{}L\"asker R., Ferrarese L., van de Ven G., Shankar F., 2014, ApJ, 780, 70
\bibitem{}Madau, P., Rees, M. J. 2001, ApJL, 551, L27
\bibitem{}Magorrian, J., Tremaine, S., Richstone, D., et al. 1998, AJ, 115, 2285
\bibitem{}Mamon G. A., Łokas E. L., 2005, MNRAS, 363, 705
\bibitem{}Marasco ,A., Cresci, G., Posti, L. et al. 2021, MNRAS, 507, 4274
\bibitem{}Marconi, A.,  Hunt, L. K. 2003, ApJ, 589, L21
\bibitem{}Marsden, C., Shankar, F., Bernardi, M. et al. 2022, MNRAS, 510, 5639
\bibitem{}Marconi, A., Risaliti, G., Gilli, R., Hunt, L.K.,Maiolino, R., Salvati, M. 2004, MNRAS, 351, 169
\bibitem{}Martin-Navarro I., Mezcua M., 2018, ApJ, 855, L20
\bibitem{}McConnell N. J., Ma C.-P., 2013, ApJ, 764, 184
\bibitem{}Menci, N., Fontana, A., Giallongo, E., Salimbeni, S.  2005,ApJ, 632, 49
\bibitem{}Menci, N., Fiore, F., Puccetti, S., Cavaliere, A. 2008, ApJ, 686, 219
\bibitem{}Menci, N., Gatti, M., Fiore, F., \& Lamastra, A. 2014, A\&A, 569, A37
\bibitem{}Menci, N., Fiore, F., Bongiorno, A., Lamastra, A. 2016, A\&A, 594, A99 
\bibitem{}Menci, N., Fiore, F., Feruglio, C. et al. 2019, ApJ, 877, 7
\bibitem{}Mo, H.J, Mao S., \& White, S.D.M., 1998, MNRAS, 295, 319
\bibitem{} Morabito, L.~K. \& Dai, X.\ 2012, ApJ, 757, 172
\bibitem{}Mowla, A.A. et al. 2019, ApJ, 880, 57
\bibitem{}Naab, T., Ostriker, J. 2017, ARAA, 55, 59
\bibitem{}Peacock, J.A. 1983, MNRAS, 202, 615 
\bibitem{}Peebles, P. J. E., 1972, Gen. Relativ. Gravit., 3, 63
\bibitem{}Peterson, B. M., 1993, PASP, 105, 247
\bibitem{}Pizzella, A., Corsini, E. M., Dalla Bonta', E., Sarzi, M., Coccato, L., Bertola, F., 2005, ApJ, 631, 785
\bibitem{}Powell, M.~C., Allen, S.~W., Caglar, T., et al.\ 2022, ApJ, 938, 77
\bibitem{}Prugniel P., Simien F., 1997, A\&A, 321, 111
\bibitem{}Reines, A. E., Volonteri, M., 2015, ApJ, 813, 82
\bibitem{}Ricci, F., Marchesi, S., Shankar, F., et al.\ 2017, MNRAS, 465, 1915
 \bibitem{}Richings, A.J., Faucher-Giguere C.-A., 2018, MNRAS, 478, 3100
\bibitem{}Robinson J. H. et al., 2021, ApJ, 912, 160
\bibitem{}Rusli, S.P. et al. 2013, ApJ, 146, 45
\bibitem{}Saglia R. P. et al., 2016, ApJ, 818, 47
\bibitem{}Sahu, N., Graham, A.W., Davies, B.L. 2019, ApJ, 876, 155
\bibitem{}Sahu, N., Graham, A.~W., \& Davis, B.~L.\ 2019, ApJ, 887, 10
\bibitem{}Sahu, N., Graham, A.~W., \& Hon, D.~S.-H.\ 2023, MNRAS, 518, 1352 
\bibitem{}Savorgnan G. A. D., Graham A. W., 2016, ApJS, 222, 10
\bibitem{}Savorgnan G. A. D., Graham A. W., Marconi A., Sani E., 2016, ApJ, 817
\bibitem{}Shankar, F., Lapi, A., Salucci, P., et al. 2006, 643, 14
\bibitem{}Shankar, F. \& Mathur, S.\ 2007, ApJ, 660, 1051
\bibitem{}Shankar, F. 2009, New Astron. Rev., 53, 57
\bibitem{}Shankar, F. et al. 2014, MNRAS, 439, 3189
\bibitem{}Shankar, F. et al. 2016, MNRAS, 460, 3119
\bibitem{}Shankar, F., Bernardi, M., Sheth, R.K. 2017, MNRAS,  466, 4029
\bibitem{}Shankar, F., Bernardi, M., Richardson, K., et al. 2019, MNRAS, 485, 1278
\bibitem{}Shankar, F., Weinberg, D.H., Marsden, C., et al. 2020a, MNRAS, 493, 1500
\bibitem{}Shankar, F. et al. 2020b, Nat. As., 4, 282
\bibitem{}Shen, X., Hopkins, P.H., Faucher-Giguere, C.-A. et al. 2020
\bibitem{}Sijacki, D., Vogelsberger, M., Genel, S., et al. 2015, MNRAS, 452, 575
\bibitem{}Silk, J. \& Rees, M.~J.\ 1998, A\&A, 331, L1
\bibitem{}Silk, J. \& Nusser, A., 2010, ApJ, 725, 556
\bibitem{}Smith, M. D. et al., 2021, MNRAS, 500, 1933
\bibitem{}Somerville, RS, Dav\'e R. 2015, ARAA, 53, 51
\bibitem{}Steinborn, L. K., Dolag, K., Hirschmann, M., Prieto, M. A.,  Remus, R.-S. 2015, MNRAS, 448, 1504
\bibitem{}Terzic, B., Graham, W. 2005, MNRAS, 197, 212
\bibitem{}Thomas, N., Davé, R., Anglés-Alcázar, D., Jarvis, M. 2019, MNRAS, 487, 5764
\bibitem{}van den Bosch R. C. E., 2016, ApJ, 831, 134
\bibitem{}van der Kruit, P.C., Freeman, K. C. 2011, ARA\&A, 49, 301
\bibitem{}van der Wel A. et al., 2014, ApJ, 788, 28
\bibitem{}Vittorini, V., Shankar, F., Cavaliere, A. 2005, MNRAS, 363, 1376
\bibitem{}Vogelsberger, M., Genel, S., Springel, V., et al. 2014, MNRAS, 444, 1518
\bibitem{}Volonteri M., Natarajan P., G\"ultekin K., 2011, ApJ, 737, 50
\bibitem{}White S. D. M., Rees M. J., 1978, MNRAS, 183, 341
\bibitem{}Yu, Q., Tremaine, S., 2002, MNRAS, 335, 965
\bibitem{}Zahid, H.J., Geller, M.J., Fabricant, D.G., Hwang, H.S. 2016, ApJ, 832, 203
\bibitem{}Zanisi, L., Shankar, F., Lapi, A. et al. 2020, MNRAS, 492, 1671
\bibitem{}Zubovas K., King A., 2012, ApJ, 745, L34

\end{thebibliography}
\end{document}